\documentclass[12pt]{article}

\usepackage[pdftex]{graphicx}

\usepackage{epsfig,cite}

\usepackage {amsmath}

\usepackage{color}
\usepackage{rotating}
\usepackage{amssymb}

\usepackage{slashed}

\include{epsf}

\newcount\timecount

\newcount\hours \newcount\minutes  \newcount\temp \newcount\pmhours

\hours = \time

\divide\hours by 60

\temp = \hours

\multiply\temp by 60

\minutes = \time

\advance\minutes by -\temp

\def\hour{\the\hours}

\def\minute{\ifnum\minutes<10 0\the\minutes

            \else\the\minutes\fi}

\def\clock{

\ifnum\hours=0 12:\minute\ AM

\else\ifnum\hours<12 \hour:\minute\ AM

      \else\ifnum\hours=12 12:\minute\ PM

            \else\ifnum\hours>12

                 \pmhours=\hours

                 \advance\pmhours by -12

                 \the\pmhours:\minute\ PM

                 \fi

            \fi

      \fi

\fi

}

\def\monthname{\relax\ifcase\month 0/\or January\or February\or

   March\or April\or May\or June\or July\or August\or September\or

   October\or November\or December\else\number\month/\fi}

\def\bold#1{\setbox0=\hbox{$#1$}%

     \kern-.025em\copy0\kern-\wd0

     \kern.05em\copy0\kern-\wd0

     \kern-.025em\raise.0433em\box0 }

\def\beq{\begin{equation}}

\def\eeq{\end{equation}}

\def\ga{\mathrel{\raise.3ex\hbox{$>$\kern-.75em\lower1ex\hbox{$\sim$}}}}

\def\la{\mathrel{\raise.3ex\hbox{$<$\kern-.75em\lower1ex\hbox{$\sim$}}}}

\def\gev{{\rm \, Ge\kern-0.125em V}}

\def\tev{{\rm \, Te\kern-0.125em V}}

\def\gyr{{\rm \, G\kern-0.125em yr}}

\def\slash#1{\rlap{\hbox{$\mskip 1 mu /$}}#1}%

\def\gappeq{\mathrel{\rlap {\raise.5ex\hbox{$>$}}

{\lower.5ex\hbox{$\sim$}}}}

\def\lappeq{\mathrel{\rlap{\raise.5ex\hbox{$<$}}

{\lower.5ex\hbox{$\sim$}}}}

\def\Toprel#1\over#2{\mathrel{\mathop{#2}\limits^{#1}}}

\def\m12{m_{1\!/2}}

\def\bea{\begin{eqnarray}}

\def\eea{\end{eqnarray}}

\def\beqar{\begin{eqnarray}}

\def\eeqar{\end{eqnarray}}

\def\m{{\cal m}}

\begin{document}

\begin{titlepage}

\pagestyle{empty}

\baselineskip=21pt


\rightline{KCL-PH-TH/2012-44, LCTS/2012-29, CERN-PH-TH/2012-268}

\vskip 1in

\begin{center}

{\large {\bf Distinguishing `Higgs' Spin Hypotheses using $\gamma \gamma$ and $WW^\ast$ Decays}}

\end{center}

\begin{center}

\vskip 0.2in

 {\bf John~Ellis}$^{1,2}$,
 {\bf Ricky Fok}$^{3}$,
 {\bf Dae Sung Hwang}$^4$,
 {\bf Ver\'onica Sanz}$^{2,3}$,
and {\bf Tevong~You}$^{1}$

\vskip 0.2in

{\small {\it

$^1${Theoretical Particle Physics and Cosmology Group, Physics Department, \\
King's College London, London WC2R 2LS, UK}\\

$^2${TH Division, Physics Department, CERN, CH-1211 Geneva 23, Swabitzerland}\\

$^3${Department of Physics and Astronomy, York University, Toronto, ON, Canada M3J 1P3}\\

$^4${Department of Physics, Sejong University, Seoul 143Ð747, South Korea}\\
}}


{\bf Abstract}

\end{center}

\baselineskip=18pt \noindent


{
The new particle $X$ recently discovered by the ATLAS and CMS Collaborations in
searches for the Higgs boson has been observed to
decay into $\gamma \gamma$, $Z Z^\ast$ and $W W^\ast$, but its spin and parity, $J^P$, remain a mystery,
with $J^P = 0^+$ and $2^+$ being open possibilities.
We use {\tt PYTHIA} and {\tt Delphes} to simulate an analysis of the angular distribution of $gg \to X \to \gamma \gamma$ 
decays in a full 2012 data set, including realistic background levels. We show that this angular distribution should provide strong
discrimination between the possibilities of spin zero and spin two with graviton-like couplings: $\sim 3 \sigma$ if a
conservative symmetric interpretation of the log-likelihood ratio (LLR) test statistic is used, and $\sim 6 \sigma$ if a less conservative
asymmetric interpretation is used. The $WW$ and $ZZ$ couplings of the Standard Model
Higgs boson and of a $2^+$ particle with graviton-like couplings are both expected to exhibit custodial symmetry. We simulate the 
present ATLAS and CMS search strategies for $X \to W W^\ast$ using {\tt PYTHIA} and {\tt Delphes}, and show that their efficiencies 
in the case of a spin-two particle with graviton-like couplings are a factor $\simeq 1.9$ smaller than in the spin-zero case.
On the other hand, the ratio of $X_{2^+} \to W W^\ast$ and $Z Z^\ast$ branching ratios is larger than that in the $0^+$
case by a factor $\simeq 1.3$. We find that the current ATLAS and CMS results for $X \to W W^\ast$ and $X \to Z Z^\ast$ decays
are compatible with custodial symmetry under both the spin-zero and -two hypotheses, and that the data expected to
become available during 2012 are unlikely to discriminate significantly between these possibilities.
}


\vfill

\leftline{
October 2012}

\end{titlepage}

\baselineskip=18pt


\section{Introduction and Summary}

A new particle $X$ with mass $\sim 125$ to 126~GeV has been discovered by the ATLAS~\cite{ATLASICHEP2012} and CMS~\cite{CMSICHEP2012}
Collaborations during their searches for the Higgs boson of the Standard Model. At first sight,
the new particle $X$ is observed to have similar characteristics to the long-sought Higgs particle $H$: it
is a boson that does not have spin one, and hence it is different in character from
gauge bosons. But is it really a (the) Higgs boson?

Answering this question will require a number of consistency checks. For example, are the $X$
couplings to other particles (at least approximately) proportional to their masses? A first step
towards answering this question was taken in~\cite{EY2}  (see also~\cite{postdiscovery}), where it was shown that the data
published in the ATLAS and CMS discovery papers~\cite{ATLASICHEP2012,CMSICHEP2012} and other public documents,
in combination with results from the TeVatron experiments CDF and D0~\cite{TevatronJulySearch}, are
inconsistent with mass-independent couplings to the $t, Z^0, W^\pm$ and $b$, and in fact
highly consistent with linear scaling $\sim M/v$, where $v \sim 246$~GeV is the expected
electroweak symmetry-breaking scale. This type of consistency check will be improved
significantly with upcoming data.

However, an even more basic question is whether the recently-discovered Higgs candidate has the
spin-parity $J^P = 0^+$ expected for the scalar Higgs boson in the Standard Model, and several ways
to test this have been proposed. For example, some of us have recently pointed out~\cite{EHSY} that the $X V$ invariant
mass distributions in associated production of the $X$ particle together with a vector boson
$V = Z$ or $W$ are in theory very different for the $J^P$ assignments $0^+, 0^-$ and $2^+$ for the $X$
particle (where we assume graviton-like couplings in the last case). We have also shown that these
differences in the $VX$ mass distributions are maintained in simulations with realistic detector cuts applied,
and hence may be used to obtain indications on the $J^P$ of the $X$ particle, if the experimental
backgrounds can be suppressed sufficiently. At the time of writing, the TeVatron experiments CDF and D0 have reported evidence for $X$
production in association with vector bosons $V$, followed by $X \to {\bar b} b$
decay, at a rate compatible with the Standard Model~\cite{TevatronJulySearch}, but have not yet reported the $VX$ mass distributions.
The most important backgrounds are expected to have small invariant masses.
(Non-)observation of a similar $VX$ signal in ATLAS or CMS at the Standard Model level with
an invariant mass distribution corresponding to the $2^+$ prediction would provide
evidence for (against) the $2^+$ hypothesis.

Here we first study the potential discriminating power of the
angular distribution of $gg \to X \to \gamma \gamma$ events, which we
simulate using {\tt PYTHIA} and the {\tt Delphes}, including realistic background levels.
We find that the data already available may be able to offer some discrimination,
and that the data likely to be available to the ATLAS and CMS
Collaborations by the end of 2012 should be able to distinguish between the
spin-zero and graviton-like spin-two hypotheses at about the 3-$\sigma$ level if a
conservative symmetric interpretation of the LLR test statistic is used, or over 6 $\sigma$ if a less
conservative asymmetric interpretation is used.

There have been many suggestions to exploit correlations among decay products of the `Higgs'
to identify its spin and parity~\cite{zz}. Some of these correlations have been incorporated in the 
Higgs search strategies adopted by the ATLAS, CMS, CDF and D0 collaborations. These are
based on the assumption that its $J^P = 0^+$ and have, in general,
lower efficiencies for detecting a particle with different $J^P$. For example, the ATLAS and
CMS searches for $H \to W W^* \to \ell^+ \ell^- \nu {\bar \nu}$ decay make use of the kinematic correlations
expected for a scalar particle, and the search by CMS for $H \to Z Z^* \to 4 \ell^\pm$ also exploits the
correlations expected in decays of a spin-zero particle.

We study here whether the logic can be inverted to
argue that the results of these searches already favour the spin-zero hypothesis for
the $X$ particle over the spin-two hypothesis. We find that the current ATLAS and CMS
measurements favour custodial symmetry for the $XWW$ and $XZZ$
both if the $X$ particle has $J^P = 0^+$ and if it has $J^P = 2^+$ and graviton-like couplings.
This result is based on simulations of the ATLAS and CMS $H \to W W^*$ searches using
{\tt PYTHIA} and {\tt Delphes}, which indicate that their efficiencies are a factor
$\simeq 1.9$ lower under the spin-two hypothesis. On the other hand, this effect is partially
offset by the ratio of the $X \to WW^\ast$ and $ZZ^\ast$ branching ratios that is larger by a factor $\simeq 1.3$ in the $2^+$ case.
Extrapolation to the full expected 2012 data set suggests that it will not be able to
discriminate significantly between the $0^+$ and graviton-like $2^+$ hypotheses using the ratio
of $X \to WW^\ast$ and $ZZ^\ast$ decays.

\section{Possible spin-parity assignments}

As is well known, the fact that the new particle has been observed to decay into a pair of on-shell photons
implies that it cannot have spin one~\cite{CMSICHEP2012,ATLASICHEP2012}. Accordingly, here we consider the spin-zero and spin-two options.
In the spin-zero case, we consider the pseudoscalar possibility $J^P = 0^-$ as an alternative to the assignment $0^+$
expected for the Standard Model Higgs boson, and we consider spin-two models with graviton-like couplings.

\subsection{Pseudoscalar couplings}

In this case, the following are the couplings to two vector bosons and a fermion-antifermion pair, respectively:
\beq
{V_\mu V_\nu}: \; \epsilon_{\mu\nu\rho\sigma}p^\rho q^\sigma, \; \; \; \; \; {\bar f}f: \; \gamma_5 \, .
\label{pseudoscalar}
\eeq
where the first term corresponds to an $\epsilon_{\mu \nu \rho \sigma} F^{\mu \nu} F^{\rho \sigma} X$ interaction term, 
with $p_\mu$ and $q_\nu$ the sum and difference of the four-momenta of the two vector bosons, respectively.
In this case, the forms of the vertices are unique, though the normalizations are arbitrary. We
assume a custodial symmetry so that the pseudoscalar couplings to $W^\pm$ and $Z$ are equal,
but make no assumptions about the ratio of the pseudoscalar couplings to the photon and gluon.

\subsection{Tensor couplings}

Several forms are possible for the couplings of a spin-two particle to two vector bosons.
It was shown in~\cite{us-G} that Lorentz and Standard Model gauge symmetries forbid
couplings of a massive spin-2 particle to two Standard Model particles through dimension-four terms in the Lagrangian and,
assuming the flavour and CP symmetries of the Standard Model, it should couple 
flavour-diagonally via dimension-five terms that take the same forms as their
energy-momentum tensors, namely
\begin{equation}
{\cal L}_{int}=-\frac{c_{i}}{M_{eff}} G^{\mu \nu} T^{i}_{\mu \nu} \, , \label{LG}
\end{equation}
where the $T^{i}_{\mu \nu}$ are the four-dimensional stress tensors of the Standard Model species $i=b, f, V, ...$,
where $V$ denotes a generic gauge boson. In scenarios with extra dimensions, 
$M_{eff}$ is the effective Planck mass suppressing the interactions ($M_{eff} \simeq$ {\cal O}(TeV)), 
whereas in composite models it denotes a scale related to confinement.

In the specific cases of vector boson fields $V_\mu$, one has:
\beq
T^{V}_{\mu \nu} \; = \;  -F_{\mu}^{\rho} F_{ \rho \nu} +  (\mu \leftrightarrow \nu) - m_V^2  V_{\mu} V_{\nu}  \, ,
\label{TV}
\eeq
where $F_{\mu \nu}$ is the field strength tensor for $V_\mu$ and the vector boson mass term $\propto m_V^2$ would be
absent for the photon and gluon.
In the case of a fermion $f$, one has
\beq
T^{f}_{\mu \nu} \; \supset \; \frac{i}{2} \, \bar{\psi} \gamma_{\mu}\partial_{\nu}\psi + (\mu \leftrightarrow \nu)+ \textrm{ h.c.} - m_f {\bar \psi} \psi g_{\mu \nu} \, .
\label{Tf}
\eeq

Since the couplings of a composite spin-two particle and a massive graviton  
Standard Model particles would both take the forms (\ref{LG}), the model dependence would appear in the coefficients $c_i$. In the case of a resonance in a strong sector, these coefficients
would reflect the underlying dynamics and the quantum numbers of the constituent
fields. For example, if the constituents do not carry color, the coupling to gluons, $c_g$, would be zero,
whereas the coupling to photons, $c_\gamma$, would reflect their electric charges.

A more specific scenario is that of massive gravitons in warped extra dimensions, 
with the Standard Model particles residing in the bulk~\cite{RSbulk}.
In five-dimensional scenarios with a factorizable metric
\beq
ds^2 \; = \; w^2(z) (\eta_{\mu \nu} dx^\mu dx^\nu - dz^2) \, ,
\label{factorizable}
\eeq
one may consider various possibilities. For example, whereas in a flat extra dimension $w(z) = 1$,
in an AdS extra dimension $w(z) = z_{UV}/z$. If a Standard Model field lives on a brane
located at $z_\ast \in (z_{UV},z_{IR})$, one has
\beq
c \; \simeq \; \frac{w(z_{IR})}{w(z_{\ast})} \, .
\label{astUVratio}
\eeq
In flat extra dimensions there would be no parametric suppression, $w = 1$ and the couplings (\ref{astUVratio})
would be universal~\footnote{We note that in flat extra dimensions one typically finds conservation of a
Kaluza-Klein parity that would forbid the coupling of the massive graviton to two light Standard
Model particles.}. However, in warped extra dimensions with $w(z_{IR}) \ll w(z_{UV})$ and for a field living
on the UV brane $c =  w(z_{IR})/w(z_{UV}) \simeq 1~{\rm TeV}/M_P$~\cite{Randall:2002tg}.
On the other hand, couplings to massless gauge bosons would be suppressed by the effective
volume of the extra dimension, namely~\cite{gap-metrics}
\beq
c \; \simeq \; 1/\int_{z_{UV}}^{z_{IR}} w(z) dz \; .
\label{cint}
\eeq
In the AdS case, this suppression would be by $\log (z_{IR}/z_{UV}) \simeq \log(M_P/{\rm TeV}) \sim 30$,
whereas in flat space the suppression would be by the entire volume of the extra dimension.
In either case, the couplings for gluons and photons would be equal: $c_g = c_\gamma$, assuming
that localized kinetic terms are not unnaturally big.

We note that, in either case, custodial symmetry is ensured for
the spin-two couplings to the massive vector bosons of the Standard Model. 
The reason is gauge invariance: the graviton couples to the gauge eigenstates $W^a$ universally, 
which implies that   $c_W = c_Z$ as $g'\to 0$. Once electroweak symmetry breaking
occurs, the graviton would feel the effect through couplings like 
$G_{\mu\nu} D^{\mu} \Sigma^{\dagger} D^{\nu} \Sigma $, where $\Sigma$ is a  physical or spurion field,
and $\langle\Sigma\rangle = v$. As $\Sigma$ should respect an approximate custodial symmetry 
(as indicated by the small value of the $T$ and $\Delta \rho$ parameters~\cite{pdg}), 
the spin-two particle inherits the approximate custodial symmetry. 

If we assume that electroweak symmetry is broken by boundary conditions on the IR brane, we
expect that the support of the $W$ and $Z$ wave functions will be suppressed in the neighbourhood
of this brane, so that $c_{W_t,Z_t} < c_{\gamma, g}$, where the $V_t$ are the transverse components. On the other hand, the longitudinal $W$ and $Z$ are localized near the IR brane~\cite{lisa-liam}, as  the wave functions of
massive fermions such as the $b$ and $t$,
so that $c_{W_L,Z_L}$,  $c_{b, t} > c_{\gamma, g}$. and the wave functions of light fermions such as the $u$ and $d$
are expected to to be concentrated closer to the UV brane, so that $c_{u, d} \ll c_{\gamma, g}$. 

To summarize, in warped extra dimensions one expects the following qualitative behaviour
\begin{eqnarray}
{\rm Warped~AdS} & : & \; c_b,  \; c_t  > c_W \simeq c_Z \gg  c_\gamma = c_g  \gg c_u , \; c_d \, . \label{AdS}
\end{eqnarray}
with, e.g., $c_b \simeq 30 \, c_{\gamma, g}$ in the case of a Randall-Sundrum model
with the third generation located near the UV brane~\cite{RSbulk,Grosmann-ferm}.
In Section~4 below, we focus on the question whether the custodial symmetry relation
$ c_W \simeq c_Z$ is compatible with the present experimental data.

\section{Angular distributions in $X_{0,2} \to \gamma \gamma$ decay}



In the case of a spin-0 particle, $X_0$, its decay products  are isotropically distributed over a 
two-dimensional sphere, so one expects a flat distribution as a function of $\cos\theta^*$, 
where $\theta^*$ is the angle of the photon relative to the beam axis in the $X$ rest frame. 
On the other hand, the $\gamma\gamma$ angular distribution will in general be non-isotropic
in the case of a spin-2 particle, $X_2$. Assuming that the gluon-gluon fusion process
dominates $X_2$ production, and graviton-like couplings of $X_2$ to both gluons and photons as discussed in Section~2, the 
$\gamma \gamma$ angular distribution in the $X_2$ centre of mass in gluon-gluon collisions was calculated in~\cite{Gao} 
(see also~\cite{DaeSung}), and is given by
\begin{equation}
	\frac{d\sigma}{d\Omega} \; \sim \; \frac{1}{4} + \frac{3}{2}\cos^2{\theta^*} + \frac{1}{4}\cos^4{\theta^*}	\quad ,
	\label{eq:spin2angulardistribution} 
\end{equation}
which can in principle be distinguished from the isotropic distribution expected in the
$0^+$ and $0^-$ cases.

In order to see whether these theoretical distributions are distinguishable in practice,
we simulated samples of spin-0 and spin-2 production accompanied by 0, 1 or 2 jets,
followed by decay to two photons:
\bea
p p \to X_{0,2} (\to \gamma\gamma) + (0, 1, 2) \; {\rm jets}  
\eea
using {\tt Madgraph5}~\cite{MG5}, which implements graviton-like couplings in the $2^+$ case. 
Our simulation includes gluon fusion, vector boson fusion and
production in association with the top and vector bosons, in the same proportions as in the case
of a conventional Higgs boson~\footnote{To the extent that these other production mechanisms
are suppressed relative to $gg \to X$, their inclusion or omission is not important. We have checked that their inclusion
in our simulation does not affect significantly the angular distributions from $gg \to X$ alone.}.
The events are then matched using the MLM scheme in {\tt PYTHIAv6.4}~\cite{PYTHIA}, and
passed through the {\tt Delphes}~\cite{Delphes} simulation code. 

\begin{figure}[h!]
\centering
\includegraphics[scale=0.34]{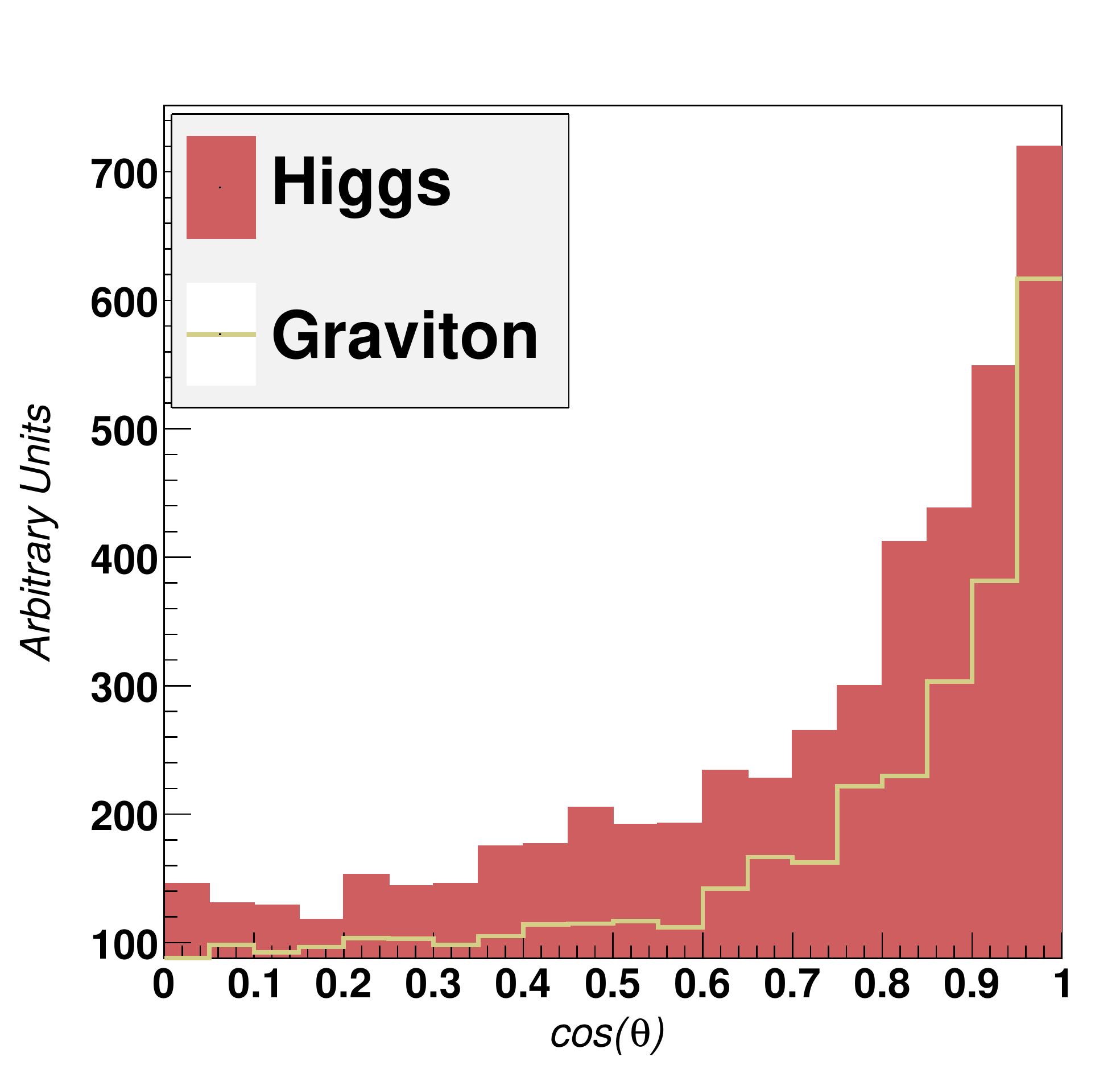}
\includegraphics[scale=0.34]{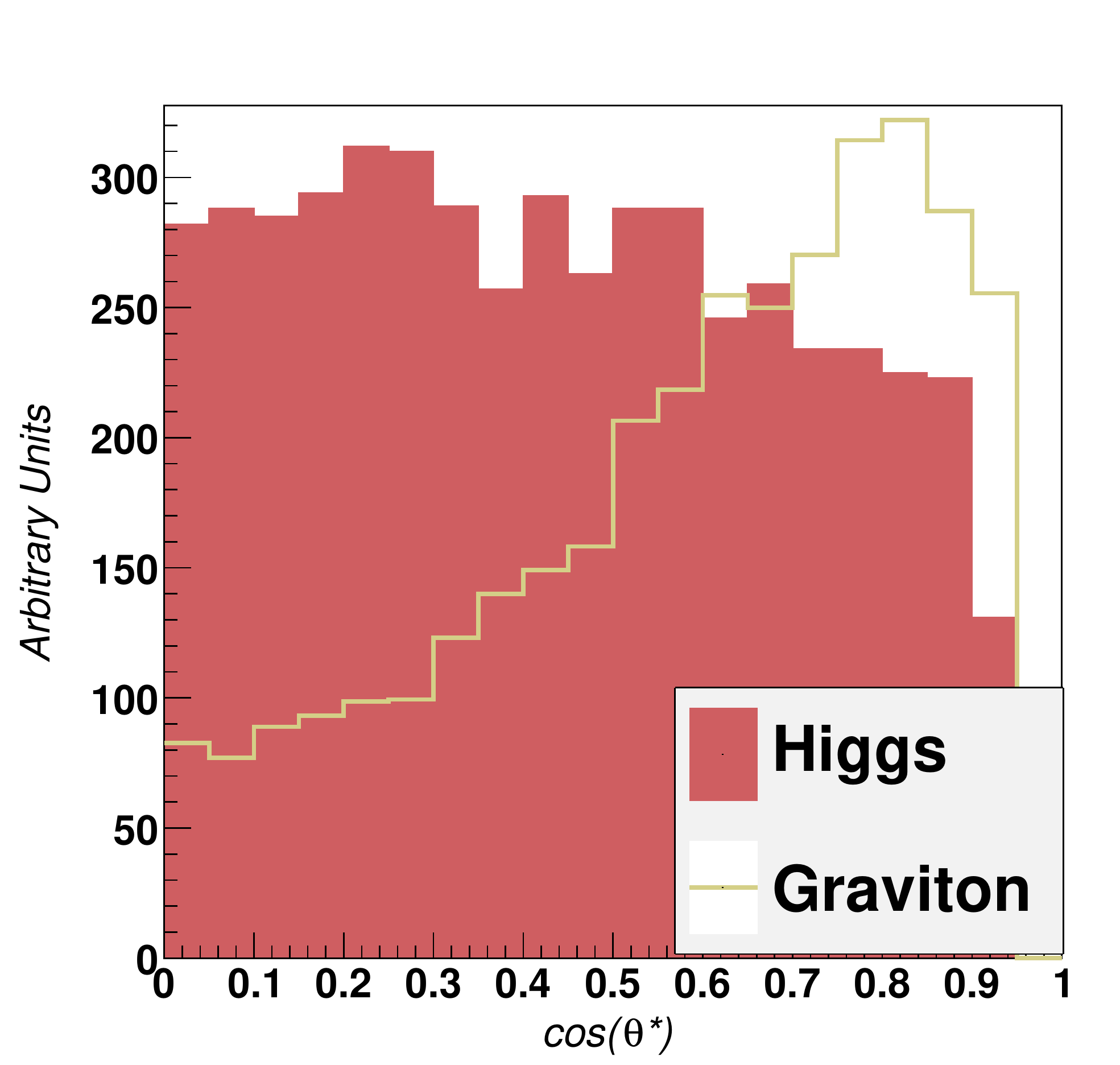}
\caption{\it Simulation of the $X_{0,2} \to \gamma \gamma$ 
polar angle distributions obtained using {\tt MadGraph5, PYTHIAv6.4}
and {\tt Delphes}, after implementing the baseline cuts
$p_T^{\gamma} > 20$~GeV and $|\eta_{\gamma}| < 2.5$. The left panel show distributions in the
laboratory frame, and the right panel show distributions in the diphoton centre-of-mass frame. 
}
\label{gamma}
\end{figure}

Figure~\ref{gamma} displays the $\cos\theta$  (left) and $\cos\theta^*$ (right) distributions after implementing
the baseline cuts $p_T^{\gamma} > 20$~GeV and $|\eta|_{\gamma} < 2.5$. We see that the theoretical difference 
between the scalar Higgs and graviton-like $2^+$ decay distributions in the rest frame of $X_{0,2}$ survives these basic cuts.

\begin{figure}[h!]
\centering
\includegraphics[scale=0.34]{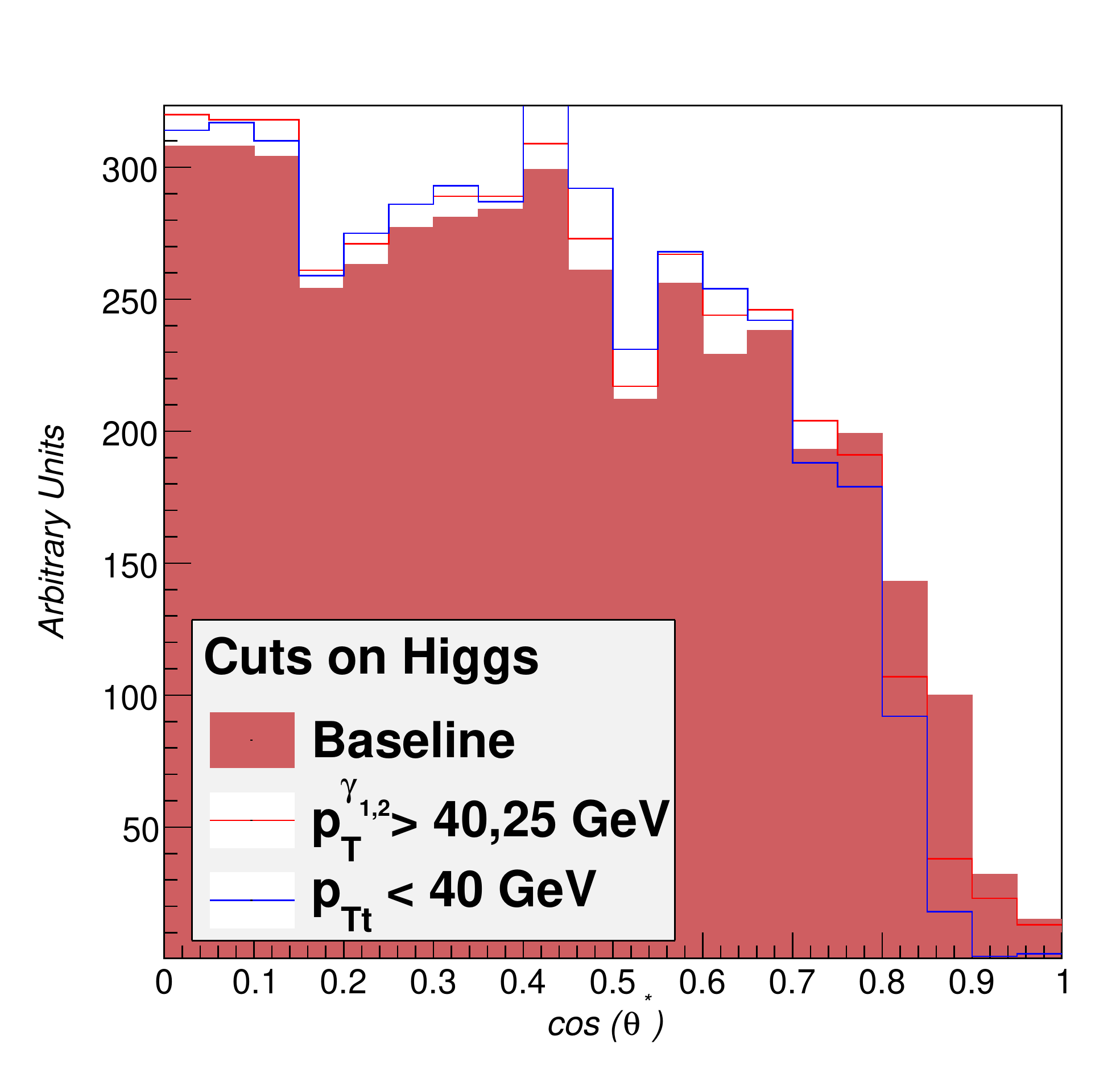}
\includegraphics[scale=0.34]{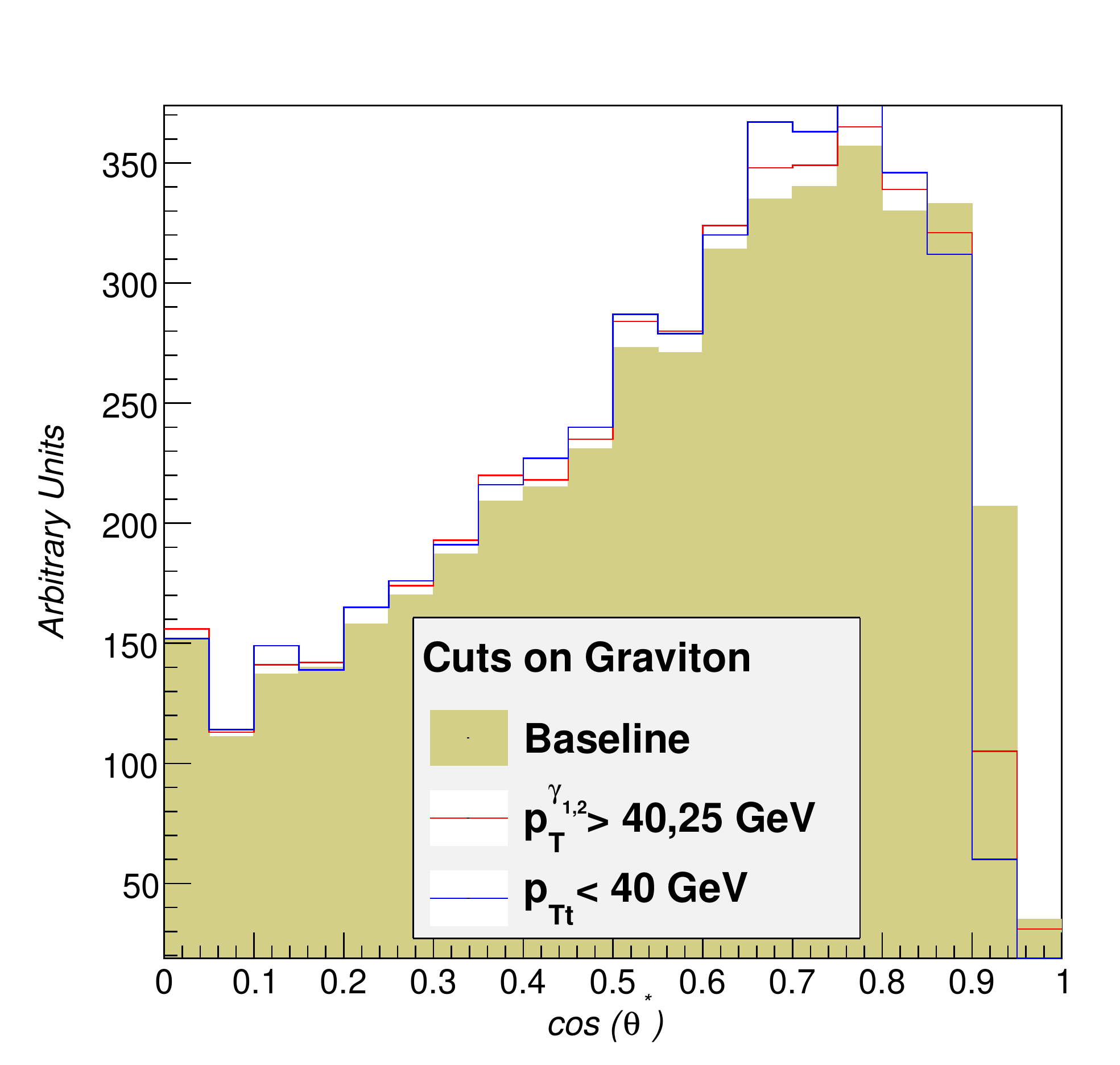}
\caption{\it The effects of different selection cuts in $p_T^{\gamma_{1,2}}$ and $p_{Tt}$ on the angular distributions
for the spin-zero case (left panel) and the graviton-like spin-two case (right panel).}
\label{effect}
\end{figure}

We have studied whether the higher-level selection cuts could affect the distributions
and the discriminating power between the spin-zero and -two hypotheses.
As shown in Fig.~\ref{effect}, we find that the distinction is quite stable under changes in the photon momentum cuts,
e.g., to $p_T^{\gamma_{1,2}}> 40, 25$~GeV, or in the $p_{Tt}$ cut
separating the glue-glue and vector-boson-fusion-enhanced processes~\footnote{We
define $p_{Tt} \equiv |\vec{p}_T^{\gamma\gamma} \times \vec{t}|$, where the thrust vector is defined as 
$\vec{t} \equiv (\vec{p}_T^{\gamma_1}-\vec{p}_T^{\gamma_1})/|\vec{p}_T^{\gamma_1}+\vec{p}_T^{\gamma_1}|$.}. 

\subsection{Toys and Statistical Procedure} 

As a first step in our analysis, we use a simple angular asymmetry variable as in~\cite{Alves} to quantify the separation significance between 
spin $0^+$ and $2^+$ as a function of the number of signal events in Monte-Carlo (MC) simulations,
starting with the idealized case of only signal events, then showing how the asymmetry can be extracted 
in the presence of background and describing how this can be simulated with toy MCs. The result is presented for 
different signal to background ($S/B$) ratios representing different event sub-categories. 
We then repeat the analysis using the complementary log-likelihood ratio (LLR) test statistic,
finding results that are somewhat more sensitive~\footnote{For a discussion of the relative merits of the LLR and 
the asymmetry variable, see~\cite{Alves}.}. 

A reference sample of 10000 spin-zero signal diphoton events from the process
$p p \to h \to \gamma \gamma$ was generated using {\tt MadGraph5 v1.4.8.3}~\cite{MG5} and passed 
through an ATLAS detector simulation based on {\tt Delphes}~\cite{Delphes}. After transforming the diphoton 
system to its centre-of-mass frame, baseline $P_T$ cuts of 40 and 25 GeV were applied on the leading 
and sub-leading photons, respectively. As shown above, the angular distribution does not vary appreciably 
before and after the detector simulation and cuts. The angular distribution of this reference sample was reweighted to obtain a spin-two reference sample. 
These reference histograms were then sampled repeatedly to provide toy histograms with numbers
of signal events that could be expected realistically. We have checked that this procedure gives results similar to generating each toy individually. 

\begin{figure}[!h]
\centerline{\includegraphics[height=7cm]{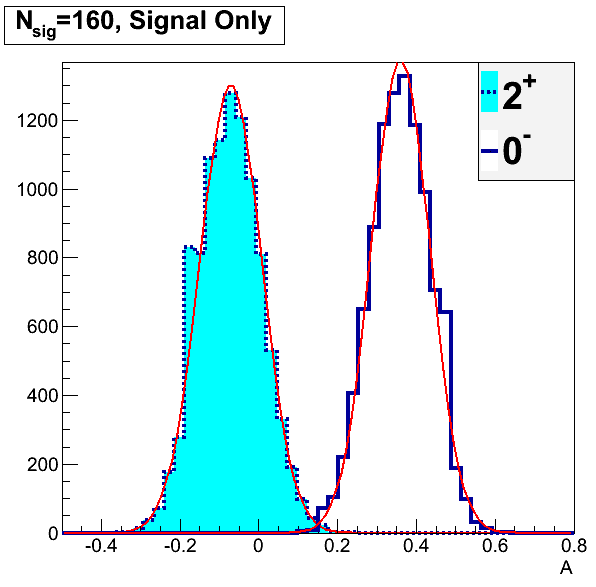}}
\caption{\it Distribution of the signal angular asymmetry variable A (\ref{eq:A}) for 10000 toys of 160 
signal events each, with a superimposed Gaussian fit in red. The histogram for the spin-zero toys is unshaded, and that for the spin-two toys
is shaded blue. These plots do not take backgrounds into account.}
\label{fig:Asigonly} 
\end{figure}

For each toy, we first quantify the shape of the distribution in $\cos{\theta^*}$ by an asymmetry variable, defined as 
\begin{equation}
	A = \frac{N_\text{centre} - N_\text{sides}}{N_\text{centre} + N_\text{sides}}	\quad , 
	\label{eq:A}
\end{equation}
where $N_\text{centre}$ is the number of events lying within the range $-0.5 \leq  \cos{\theta^*} \leq 0.5$ and 
$N_\text{sides}$ is the number of events outside this range. Populating a histogram of the asymmetry value 
for each toy gives a distribution around different means for the spin-zero and -two toys,
as illustrated in Fig. \ref{fig:Asigonly} for 10000 toys of 160 signal events.  

Using the asymmetry (\ref{eq:A}) as our test statistic, with a distribution that is fit well by a Gaussian (as seen in Fig.~\ref{fig:Asigonly})
to obtain a normalized probability distribution function $\text{pdf}(\Lambda)$, we quantify the separation 
significance using two different methods denoted as `asymmetric' and `symmetric'. 

In the {\it asymmetric} method 
one is biased towards verifying hypothesis $S_1$ and excluding hypothesis $S_2$. Thus the value of the asymmetry 
that is expected to be measured by an experiment is taken to be the mean of the distribution for $S_1$, namely
$\Lambda_{S_1}^\text{obs} \equiv \Lambda_{S_1}^\text{mean}$, and the extent to which we can exclude hypothesis 
$S_2$ is the area $\beta$ under the tail of $\text{pdf}_{S_2}(\Lambda)$: 
\begin{equation*}
	\beta = \frac{\int_{-\infty}^{\Lambda_{S_1}^\text{obs}} \text{pdf}_{S_2}(\Lambda)d\Lambda}{\int_{-\infty}^{\infty} \text{pdf}_{S_2}(\Lambda)d\Lambda}	\quad . 
\end{equation*}
For $\Lambda_{S_1}^\text{mean} > \Lambda_{S_2}^\text{mean}$ the integral is taken over the other side of the distribution,
so as to obtain instead the area under the right tail of the distribution of $S_2$. 
The quantity $\beta$ is also known as the `Type II' error, namely the probability of wrongly accepting hypothesis A
in the case that actually $S_2$ is true. 

In the {\it symmetric} method the two hypotheses are treated equally. One defines a $\Lambda_\text{cut-off}$
for which $\alpha = \beta$, namely the area $\alpha$ under the right tail of $S_1$ is equal to the area $\beta$
under the left tail of $S_2$ (in the example where $\Lambda_{S_1}^\text{mean} < \Lambda_{S_2}^\text{mean}$). 
Thus, whatever the value of $\Lambda_\text{obs}$ found by an experiment, if it lies to the left of 
$\Lambda_\text{cut-off}$ hypothesis $S_1$ is accepted and $S_2$ rejected, and vice versa if 
$\Lambda_\text{obs} > \Lambda_\text{cut-off}$. The significance is given by $\alpha=\beta$.  

Both approaches are justified in that there is strong motivation for prioritizing the spin-zero hypothesis,
and thus quoting the asymmetric significance (see also~\cite{Alves}). On the other hand, the symmetric approach
is more objective and conservative (see~\cite{Cousins}). In the following we quote results obtained using both methods. 

The significance $\alpha$ is translated into $n$ standard deviations by finding the equivalent area 
under a standard Gaussian distribution~\footnote{This is the one-sided definition most commonly used in 
the literature, as opposed to the two-sided convention sometimes seen, which generally
yields a higher number of standard deviations for the same p-value.}:
\begin{equation}
	\alpha = \frac{1}{\sqrt{2\pi}} \int_n^\infty e^{-\frac{x^2}{2}} dx	\quad . 
\label{alpha2}
\end{equation}
For example, $\alpha=0.05$ corresponds to $n=1.64$, and the discovery standard of $n=5$
corresponds to $\alpha = 2.87 \times 10^{-7}$. 

\begin{figure}[!h]
\centerline{\includegraphics[height=7cm]{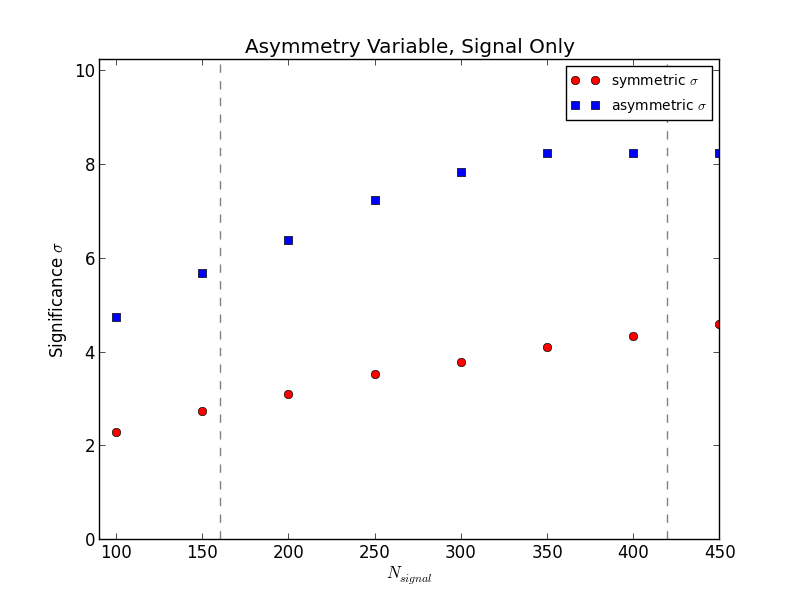}}
\caption{\it Numbers of standard deviations $\sigma$ of separation (\ref{alpha2}) for two different methods of 
interpreting the statistical significance, as functions of the numbers of signal events with no background
taken into account. The blue (red) dots are obtained using the asymmetric (symmetric) method, respectively. 
The dotted lines on the left and right are the expected yields in high and low $S/B$ categories, respectively,
with 30 $\text{fb}^{-1}$ of integrated luminosity.} 
\label{fig:AvsNsigonly} 
\end{figure}

In Fig.~\ref{fig:AvsNsigonly} we show the separation significance defined in this way for the symmetric and 
asymmetric interpretations as functions of the numbers of signal events, neglecting backgrounds. 
The vertical dotted lines indicate the expected number of signal events in two categories that 
can be reached with 30 $\text{fb}^{-1}$ of 8 TeV data collected by the end of the year. 
The expected yields in the two diphoton categories are obtained by combining the 
CMS BDT categories 0,1 and 2,3 from Table 2 of~\cite{CMSICHEP2012}. 
These correspond to high and low signal-to-background ($S/B$) ratios of approximately 0.42 and 0.19, respectively.   

Note that we have not included backgrounds in Fig.~\ref{fig:AvsNsigonly}. 
This figure should therefore be taken as an idealized limit of what could be achieved assuming a perfect
separation of the signal from the background. In the next Section we give an estimate how the background would affect this.

\subsection{Background Simulation}

A reference sample of 20K QCD continuum background $p p \to \gamma \gamma$ events
with the parton-level invariant masses of the diphoton pairs between 124 and 126 GeV was generated. The same detector simulation and cuts as above were applied. This
sample was then used to give a number of toy background events corresponding to the desired S/B ratio. 

For each toy the background and signal events are added together to give a total
sample representing the available experimental information. All that can be measured 
is the total asymmetry in the signal region, $A_\text{tot}$. However the signal asymmetry can be extracted, since 
\begin{align}
	A_\text{tot.} &= \frac{N_\text{centre} - N_\text{sides}}{N_\text{centre} + N_\text{sides}} 	\notag \\
		&= \frac{(N^s_\text{centre} + N^b_\text{centre}) - (N^s_\text{sides} + N^b_\text{sides})}{N^s_\text{centre} + N^b_\text{centre} + N^s_\text{sides} + N^b_\text{sides}}		\notag \\
		&= f A_s + (1-f) A_b		\quad ,
	\label{eq:extractedAs}
\end{align}
where we assume that $f = N_s/(N_b + N_s) = (N_\text{tot} - N_b)/N_\text{tot}$ is known, and the asymmetry of the background, $A_b$, can in principle be measured with high accuracy
in the invariant diphoton mass sideband regions outside the signal region. 

The error in the experimental determination of $f$ is the main limiting factor in reconstructing $A_s$. 
To simulate the effect of this, for each toy we randomly sample $N_b$ from a Gaussian centered 
around the true value of $N_b$, with a one-$\sigma$ width of $\sqrt{N_b}$, the statistical error. 
We assume that $N_\text{tot}$ is measured much more accurately, so that its error can be 
neglected\footnote{Adding a 1\% error in $N_\text{tot}$ does not affect the results.}. We calculate $A_b$ using the background reference sample,
so as to simulate the measurement from the sidebands that is assumed to have much higher statistics.   

\begin{figure}[!h]
\centering
\mbox
{
	\includegraphics[height=7cm]{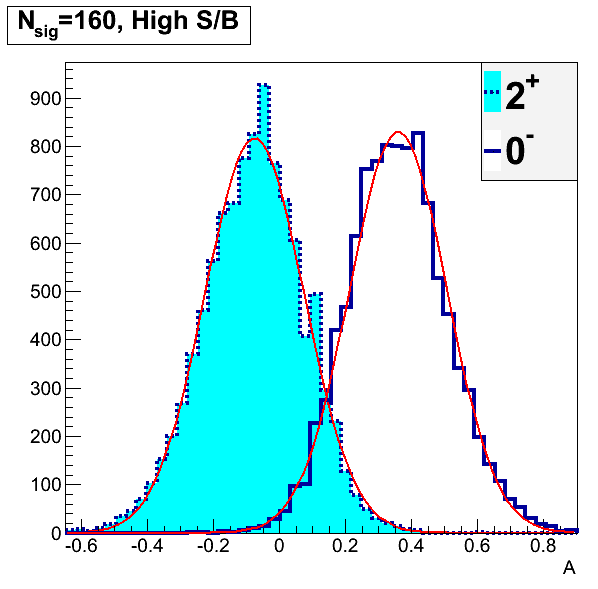}
	\includegraphics[height=7cm]{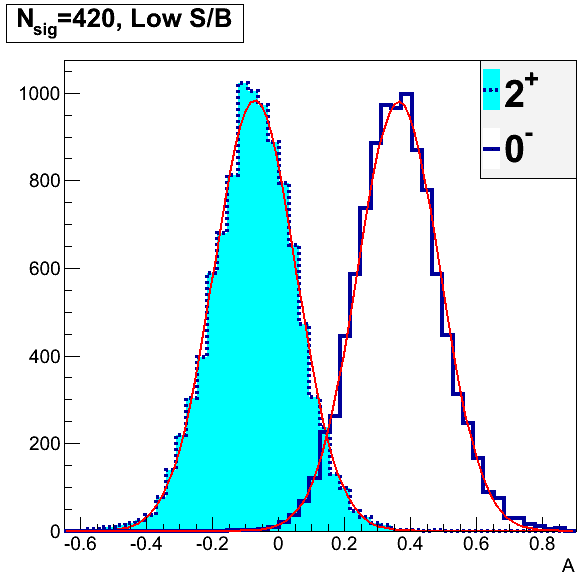}
}
\caption{\it Distribution of the extracted signal angular asymmetry variable $A$ (\ref{eq:A}) for 10000 toys each. 
The histogram for the spin-zero toys is unshaded, and that for the spin-two toys
is shaded blue. These simulations include the background with values of $S/B = 0.42$ (left) and $0.19$ (right).} 
\label{fig:Asb043} 
\end{figure}

\begin{figure}[!h]
\centering
\mbox
{
	{\includegraphics[height=6cm]{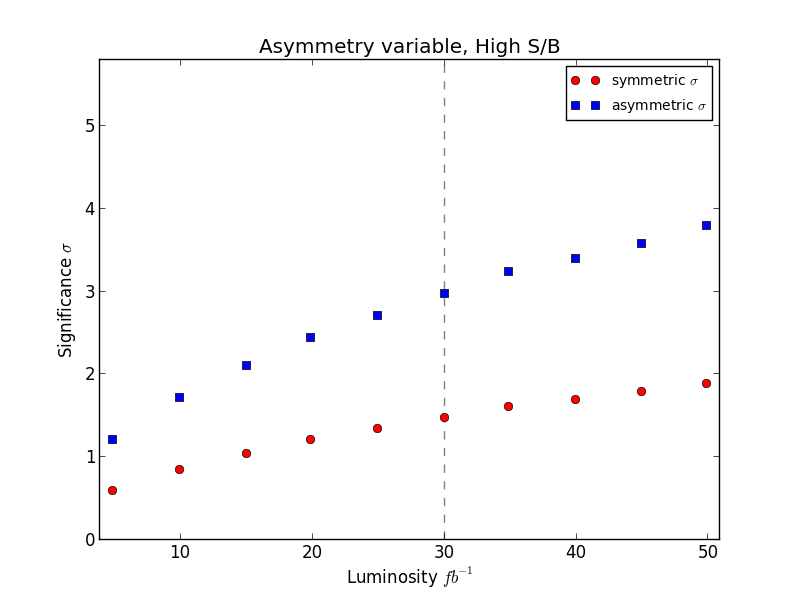}}
	{\includegraphics[height=6cm]{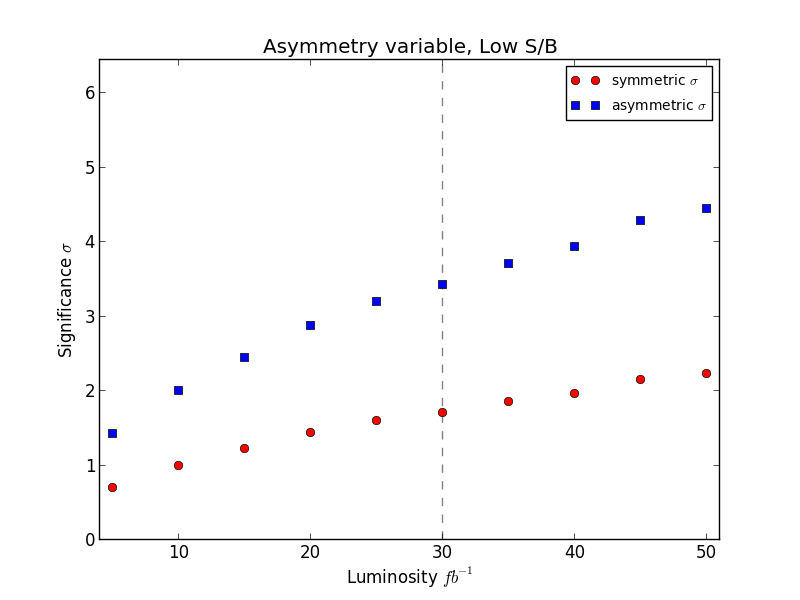}}
}
\caption{\it Numbers of standard deviations $\sigma$ of separation (\ref{alpha2}) for two different methods of 
interpreting the statistical significance of the measurement of the angular asymmetry $A$, 
as functions of luminosity with $S/B = 0.42$ (left) and $0.19$ (right). 
The blue (red) dots are obtained using the asymmetric (symmetric) method, respectively. Dotted lines
indicate the luminosity expected by the end of 2012.} 
\label{fig:AvsNwithbkg} 
\end{figure}

As mentioned above, the benchmark luminosity that the experiments hope to attain in 2012 is 30 $\text{fb}^{-1}$ at 8 TeV.
The CMS diphoton search separates events into categories that can be approximated by two samples with high and low $S/B = 0.42$ and 0.19,
respectively, with $\sim$160 and 420 expected signal events respectively for 30 $\text{fb}^{-1}$. 
The resulting distribution of $A_s$ extracted from this simulated measurement of $A_\text{tot}$, $A_b$, and $f$ per toy is 
illustrated in Fig. \ref{fig:Asb043} for 160 signal events in the high S/B category. 
Fig. \ref{fig:AvsNwithbkg} shows the separation significance as a function of luminosity for the two categories. 

We see from Fig. \ref{fig:AvsNwithbkg} that these simulations translate into a significance of 
3 to 3.5 $\sigma$ in the asymmetric interpretation for each category. Ideally, a more detailed simulation
should be done in which the events for each category are output from the BDT that sorts them, 
as this may affect the angular distribution more significantly than the simple cuts used here. 
As a basic check, we have verified that placing $|\eta < 1.4|$ cuts on both photons to simulate barrel-barrel events does not alter our results substantially. 

\subsection{Log-Likelihood Ratio Test Statistic}

An alternative test statistic is the log-likelihood ratio (LLR), for which the likelihood 
$\mathcal{L}_S$ for spin hypothesis $S$ in a single toy pseudo-experiment is defined as
\begin{equation*}
\mathcal{L}_S = \prod_i^\text{events} \text{pdf}_{S}(x_i) 	\quad .
\end{equation*}
The probability density function in $\cos{\theta^*}$ of the signal is extracted from the probability 
to lie in a bin of the high-statistics MC from which the toys are sampled. After calculating the 
likelihoods for both hypothesis $S_1$ and $S_2$ we obtain the LLR for that toy by taking $-2\ln{\frac{\mathcal{L}_{S_1}}{\mathcal{L}_{S_2}}}$. 
Thus, if we generate a set of toys for spin hypothesis $S_1$ ($S_2$), the LLR distribution will be centred around a negative (positive) mean. 
We may the quantify the separation between these two distributions as described previously.

\begin{figure}[!h]
\begin{centering}
\mbox
{
	\centerline
	{
		{\includegraphics[height=5cm]{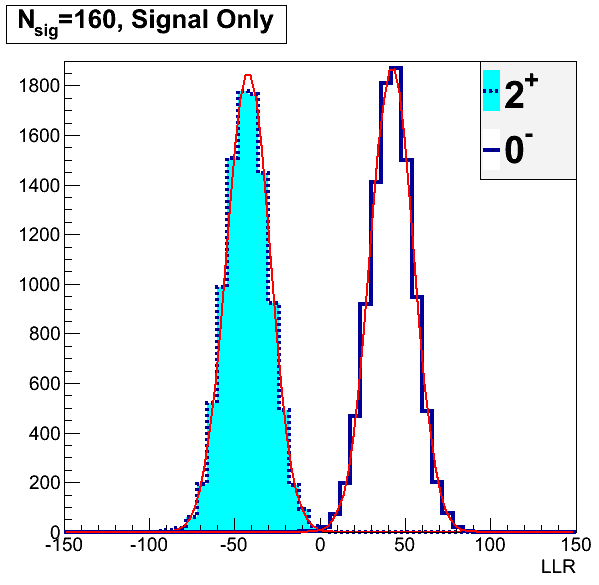}}
		{\includegraphics[height=5cm]{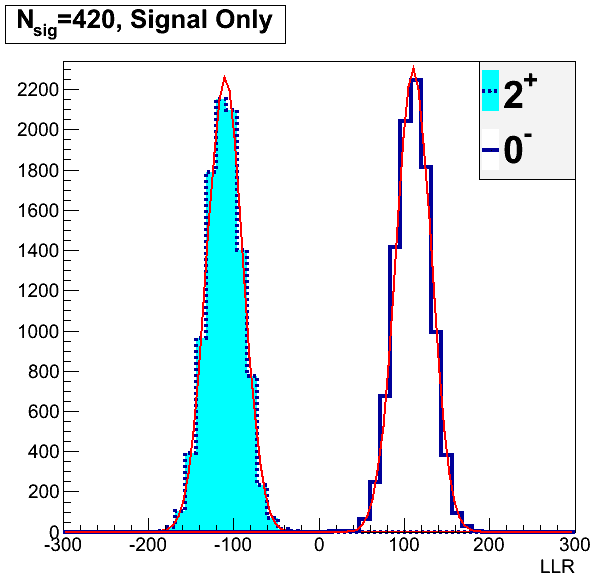}}
	}
}
\mbox
{
	\centerline
	{\includegraphics[height=6cm]{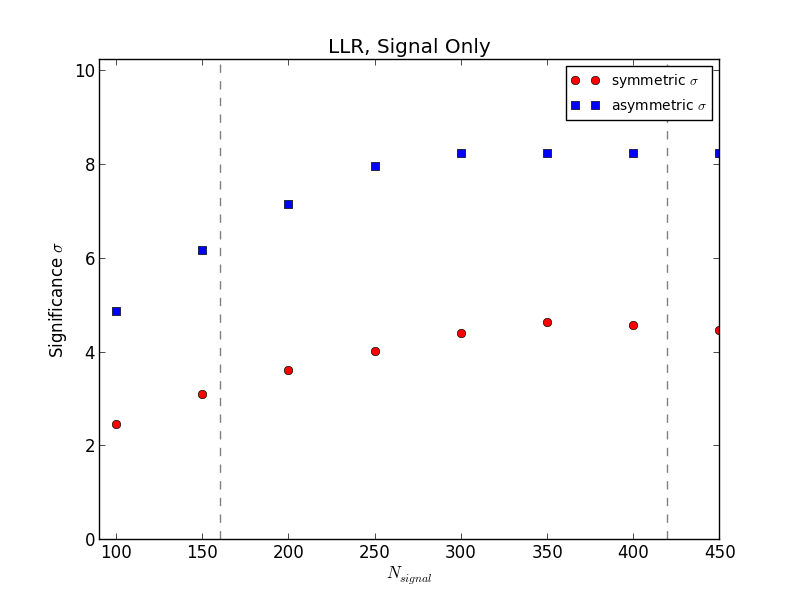}}
}
\end{centering}
\caption{\it An example distribution of the test statistic for 160 and 420 signal events, above, and the
separation significance obtained using the LLR test statistic as a function of the number of signal events, bottom.
Note that no backgrounds are included here. } 
\label{fig:LLRnobkg} 
\end{figure}

For a pure signal with no backgrounds, the LLR distributions for spin $0^+$ and $2^+$ with 160 signal 
events are shown in Fig.~\ref{fig:LLRnobkg}. Also plotted in Fig.~\ref{fig:LLRnobkg} is the separation significance 
as a function of the number of signal events, though we emphasize that this is in idealized limit in 
which the signal events can be perfectly extracted from the background. 

In order to include the effects of the background, we combine the background and signal MC to form a total MC, 
then simulate the extraction of the signal by subtracting statistically the number of background events expected
in each bin from the total number of events in that bin. The number of background events per bin is smeared 
using a Gaussian centred on the true value, with a one-$\sigma$ width given by the statistical error 
$\sqrt{N_\text{bkg}^\text{bin}}$. If the randomly smeared number of background events exceeds the total 
number of events in that bin the corresponding bin of the measured signal histogram is set to zero (since it cannot be negative). 

\begin{figure}[!h]
\centering
\mbox
{
	{\includegraphics[height=6cm]{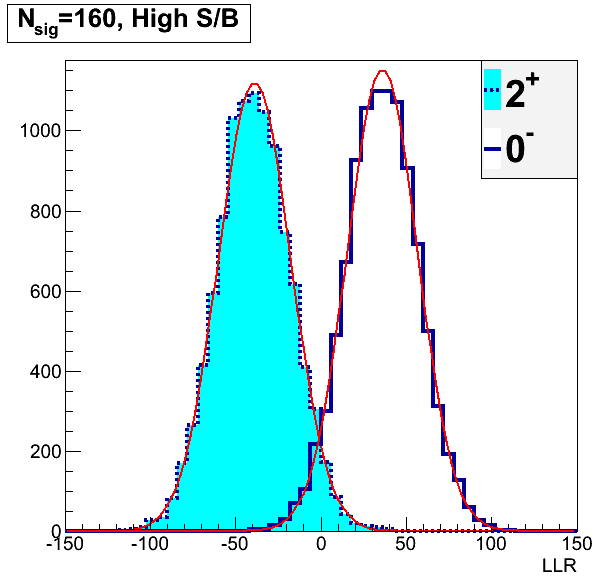}}
	{\includegraphics[height=6cm]{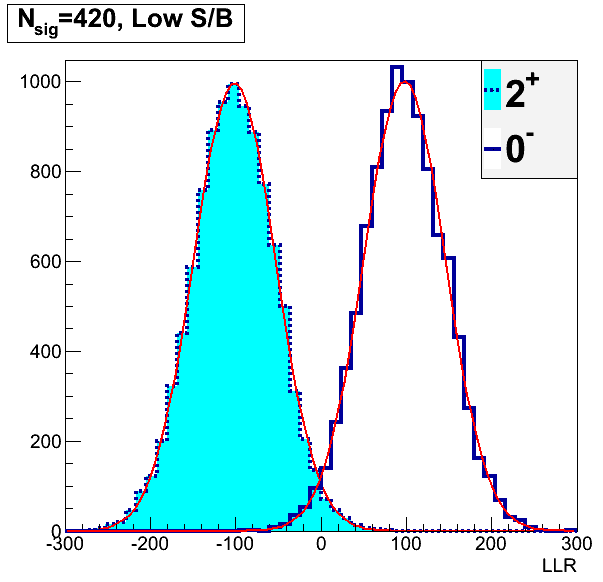}}
}
\mbox
{
	{\includegraphics[height=6cm]{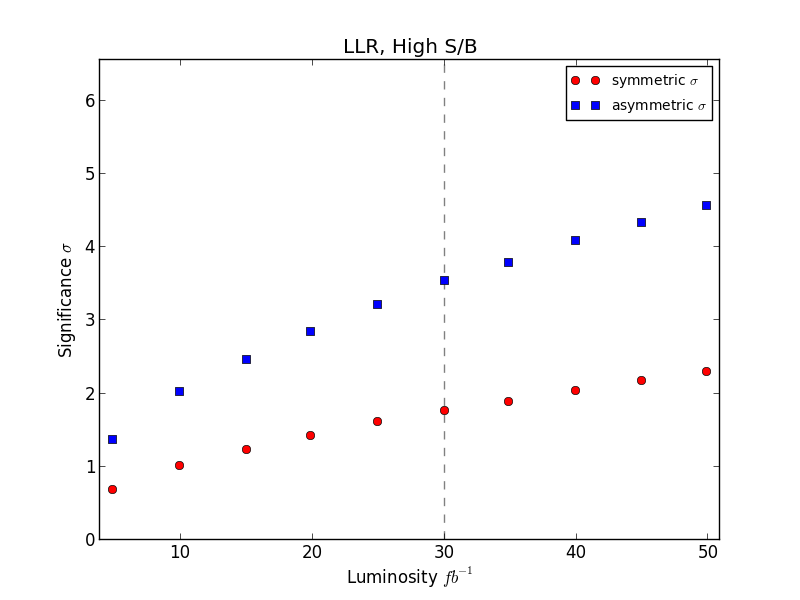}}
	{\includegraphics[height=6cm]{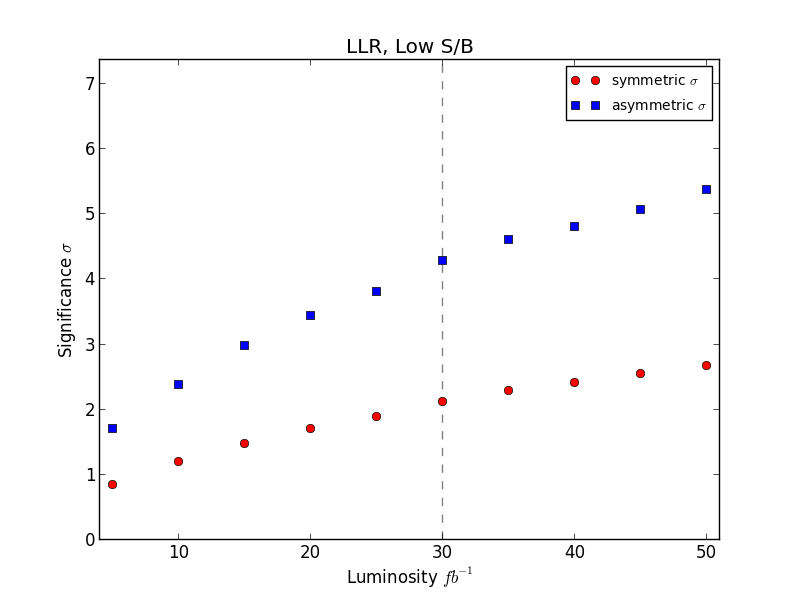}}
}
\caption{\it Separation significance using the LLR test statistic as a function of the luminosity (lower panel) 
and an example distribution of the test statistic for the 160 and 420 signal events expected in the high- and low-$S/B$
categories at 30~$\text{fb}^{-1}$ (upper panels). The backgrounds are included, 
with $S/B = 0.42$ (left) and 0.19 (right), respectively.} 
\label{fig:LLRwithbkg} 
\end{figure}

\begin{figure}[!h]
\centering
\includegraphics[height=9cm]{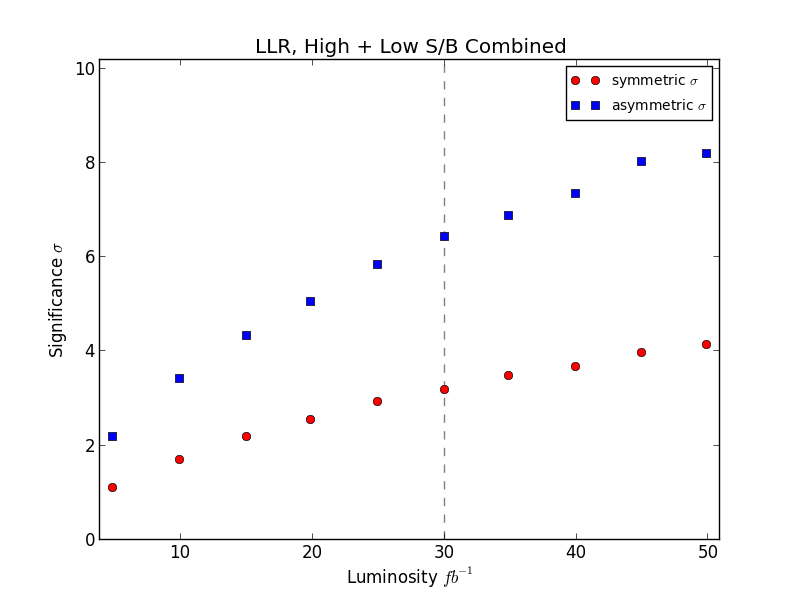}
\caption{\it Separation significance using the LLR test statistic as a function of luminosity for the combination of high- and low-$S/B$
categories. The dotted line indicates the expected reach by the end of 2012. }
\label{fig:LLRcombination}
\end{figure}

Distributions in LLR and separation significance plots similar to those in Fig.~\ref{fig:LLRnobkg}
are shown in Fig.~\ref{fig:LLRwithbkg}, with the backgrounds now taken into account, for both high and low
values of $S/B = 0.42$ and 0.19, respectively. We see that the high (low) $S/B$ category for 30 $\text{fb}^{-1}$ of 8 TeV data, 
corresponding to 160 (420)  signal events, yields a separation significance around 3.5 (4.2) $\sigma$ using the asymmetric method. 
A combination of the high and low categories can achieve over 6-$\sigma$ separation using the asymmetric
method, as is seen in Fig.~\ref{fig:LLRcombination}, and $\sim 3\, \sigma$ using the symmetric method. 

\section{Angular, mass and $m_T$ Distributions in $X_{0,2} \to WW^* \to \ell^+\nu_l \ell^-\bar{\nu}_l$ Decay} 

As discussed in~\cite{DaeSung}, the lepton momentum distributions and correlations are
very different for the $X_{0,2}$ hypotheses. In the spin-zero case, the spins of the $W^\pm$
and $W^{\mp *}$ must be {\it antiparallel}, implying that the
charged leptons $\ell^\pm$ produced in their decays appear preferentially in the {\it same}
hemisphere. On the other hand, in the spin-two case, the spins of the $W^\pm$
and $W^{\mp *}$ must be {\it parallel}, implying that their daughter $\ell^\pm$ appear
preferentially in the {\it opposite} hemisphere. As pointed out in~\cite{DaeSung}, these
differences in the decay kinematics imply that the dilepton invariant mass $m_{\ell\ell}$
is generally {\it smaller} in $X_0$ decay than in $X_2$ decay, as is the difference $\phi_{\ell\ell}$
between the $\ell^\pm$ azimuthal angles. A corollary is that the net transverse momenta
of the $\ell^+ \ell^-$ pair, $p_T^{\ell_1,\ell_2}$, is generally {\it larger} in $X_0$ decay than in $X_2$ decay.
We now address the question whether and to what extent these differences survive the event
selections and cuts made by ATLAS and CMS.

Regarding the simulation details, we created new models in {\tt Feynrules}~\cite{Feynrules}, 
including the pseudoscalar and graviton-like spin-two couplings described in Section 2. 
We then interfaced with  {\tt MadGraph5}~\cite{MG5} using the {\tt UFO} model format~\cite{UFO}.
We incorporate hadronization and showering effects using {\tt PYTHIA}~\cite{PYTHIA},
and detector effects using {\tt Delphes}~\cite{Delphes}.

\subsection{Simulation of ATLAS and CMS event selections}

\begin{figure}[h!]
\centering
\includegraphics[scale=0.34]{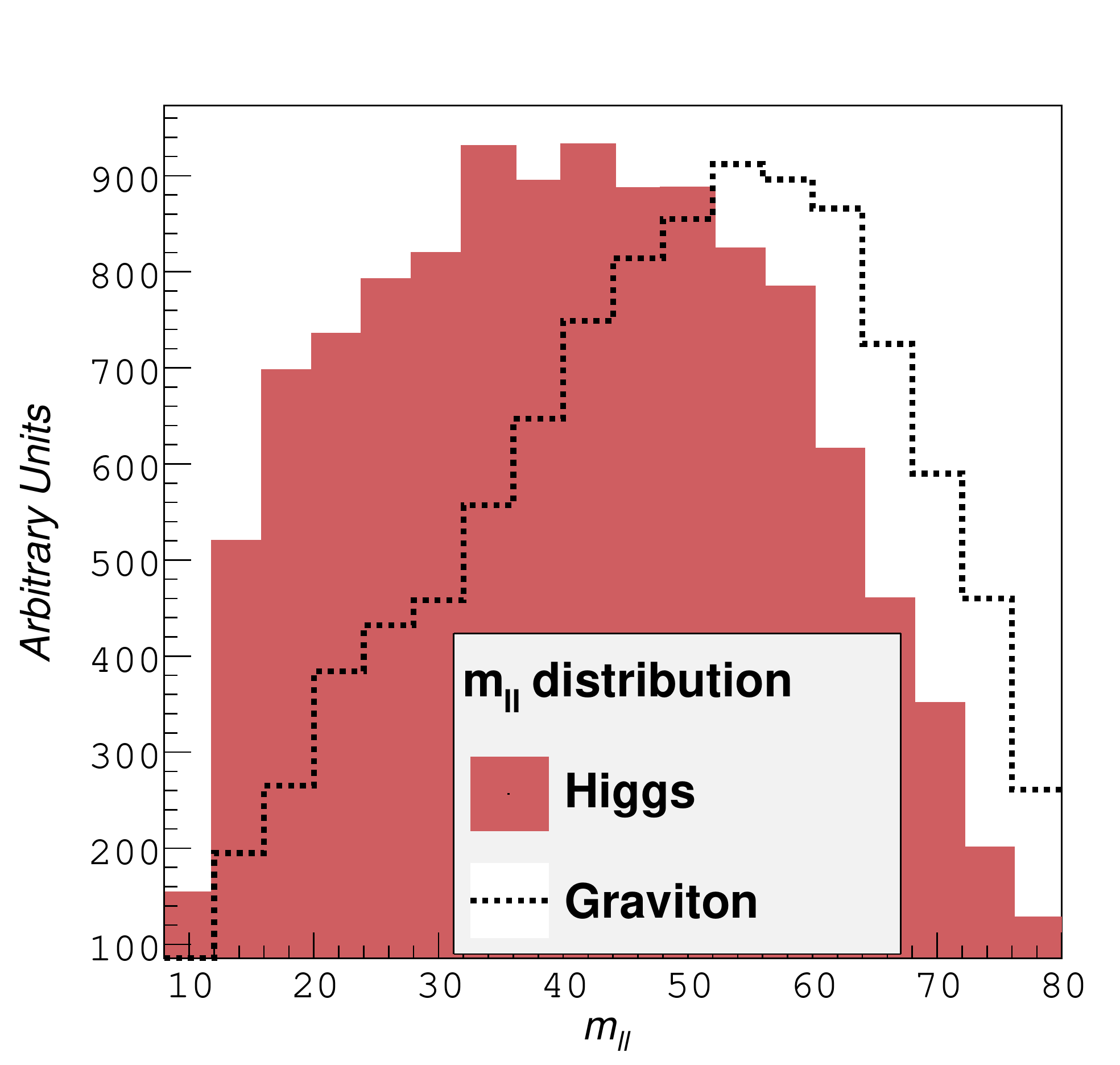}
\includegraphics[scale=0.34]{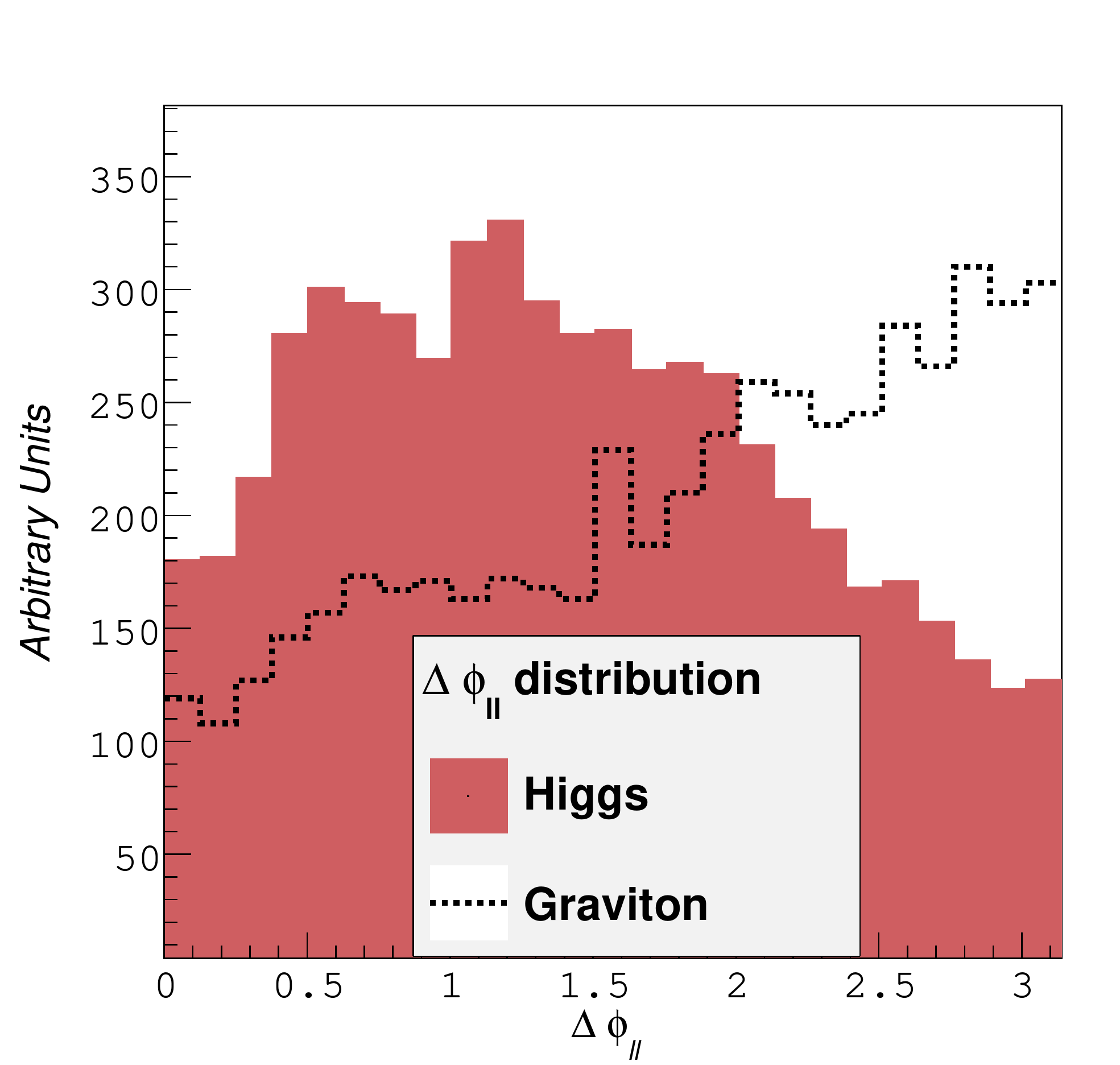}
\caption{\it Results from simulations of the $X_{0^+,2^+} \to WW^* \to \ell^+\nu_l \ell^-\bar{\nu}_l$ signals,
using {\tt PYTHIA} and {\tt Delphes} and the baseline cuts described in the text.
The left (right) panel displays the $m_{\ell\ell}$ ($\phi_{\ell\ell}$) distributions.}
\label{phill}
\end{figure}

We start by implementing the  baseline cuts of the ATLAS and
CMS analyses: two isolated leptons ($=e,\mu$) of $p_T> 15$ GeV and $|\eta|<$ 2.5.
As seen in Fig.~\ref{phill}, the distributions in the dilepton invariant mass $m_{\ell\ell}$ (left)
and in the relative azimuthal angle $\phi_{\ell\ell}$ (right) are very different in the baseline $X_{0}$
 and $X_2$ simulations. These differences reflect the
kinematical effects noted earlier. In particular, $m_{\ell\ell}$ is generally smaller in the $X_0$ case
than in the $X_2$ case, as seen in Fig.~7 of~\cite{DaeSung}, as is $\phi_{\ell\ell}$, reflecting the
angular distribution shown in Fig.~6 of~\cite{DaeSung}.

Our next step is to simulate the ATLAS search for $X \to WW^* \to \ell^+\nu_l \ell^-\bar{\nu}_l$  events.
This is based on a selection of events with two opposite-sign, unlike-flavour leptons
with $p_T^{\ell_1,\ell_2} > 25, 15$~GeV in the central region, and invariant mass $m_{\ell \ell} \in [10, 80]~{\rm GeV}$~\footnote{We
implement the relevant quality and isolation criteria at the level of the {\tt Delphes} simulation.}.
The events are then separated into categories with 0, 1 and 2 anti-$k_T$ jets (defined by cones $R=0.4$)
and $p_T>$ 25, 30 GeV in the central and forward regions, respectively.
In the 0- and 1-jet samples used here, the dilepton invariant mass upper bound is tightened to 50 GeV.
The following set of cuts is then applied to the 0-jet sample:
\bea
E^{miss}_{T,rel}> 25 \textrm{ GeV, }  p_{T}^{\ell\ell} > 30 \textrm{ GeV, and }  |\Delta \phi_{\ell\ell}| < 1.8 \ ,
\eea
where $E_T^{rel} \equiv \slash{E}_T \, \sin \Delta \phi_{min}$
with $\Delta \phi_{min} \equiv {\rm min}(\Delta \phi, \pi/2)$ and $\Delta \phi$ the minimum angle between
the missing-energy vector and the leading lepton, the subleading lepton or any jet with $p_T>$ 25 GeV.
In the 1- and 2-jet case, there is an extra cut
\bea
|\vec{p}_{T}^{tot}|=|\vec{p}_{T}^{\ell\ell}+ \vec{p}_{T}^{j}+\vec{E}_{T}^{miss}| < 30 \textrm{ GeV}
\eea
as well as a $b$-tag veto. In the 2-jet case we also implemented the vector-boson-fusion cuts
\bea
m_{jj} > 500  \textrm{ GeV and } |\Delta y_{jj}| > 3.8 \ .
\eea
Finally, a cut $m_{T} \equiv \sqrt{ (E_T^{\ell\ell}+\slash{E}_T)^2-|\vec{p}_T^{\ell\ell}+\slash{\vec{p}}_T|^2} \in [93.75, 125]$~GeV is applied to emulate the fit to the distribution performed in the ATLAS analysis.

\begin{figure}[h!]
\centering
\includegraphics[scale=0.24]{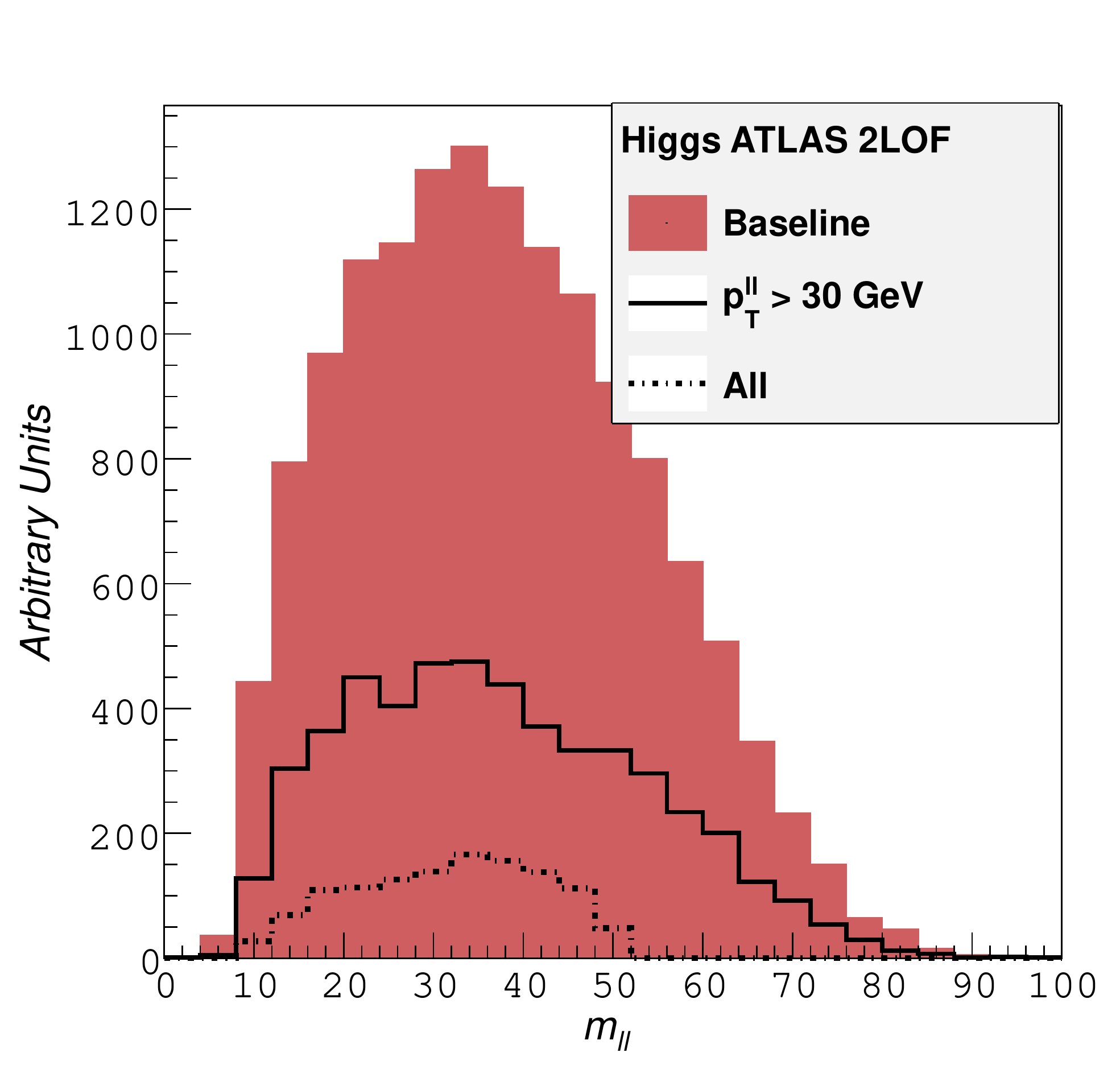}
\includegraphics[scale=0.24]{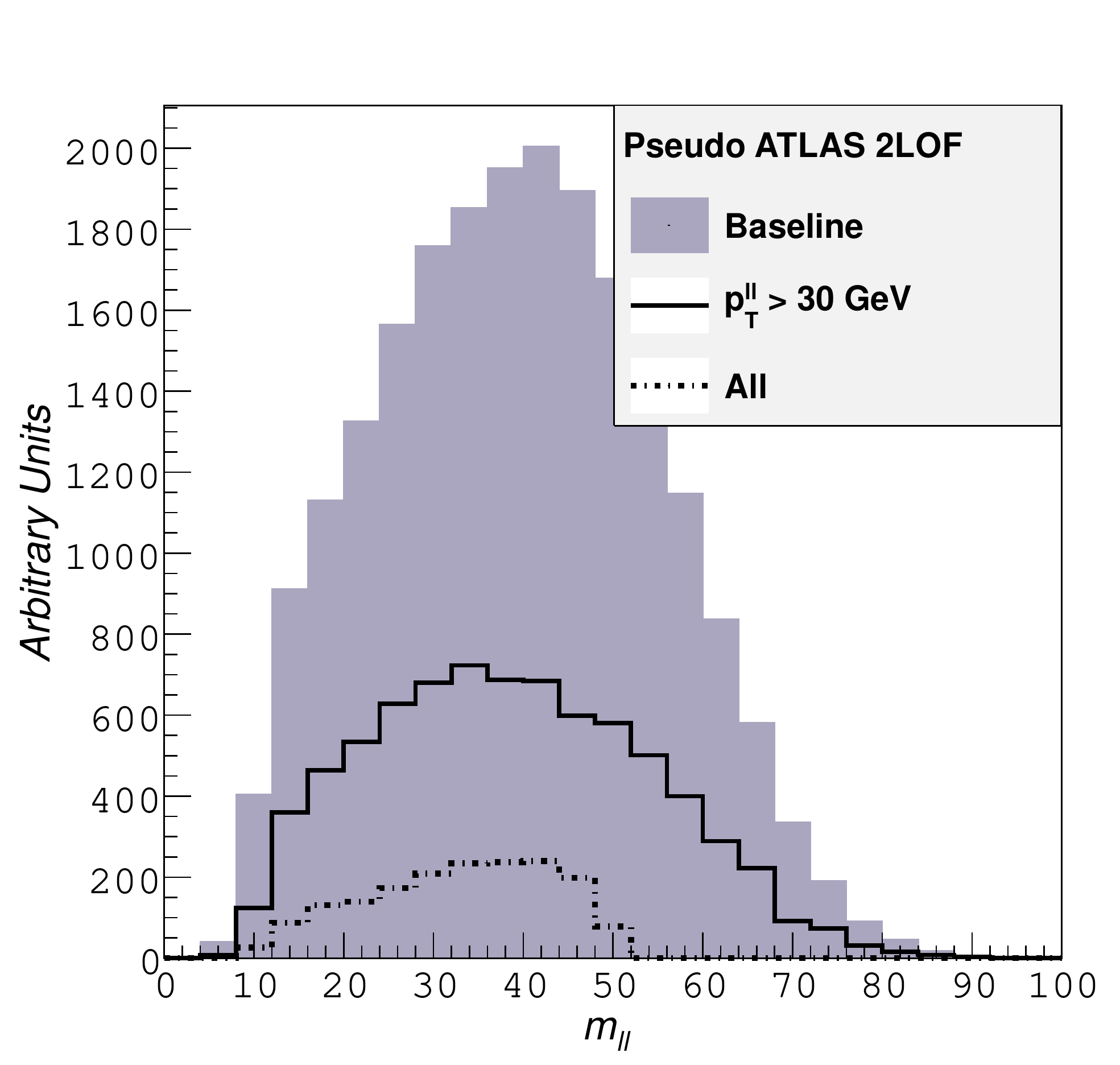}
\includegraphics[scale=0.24]{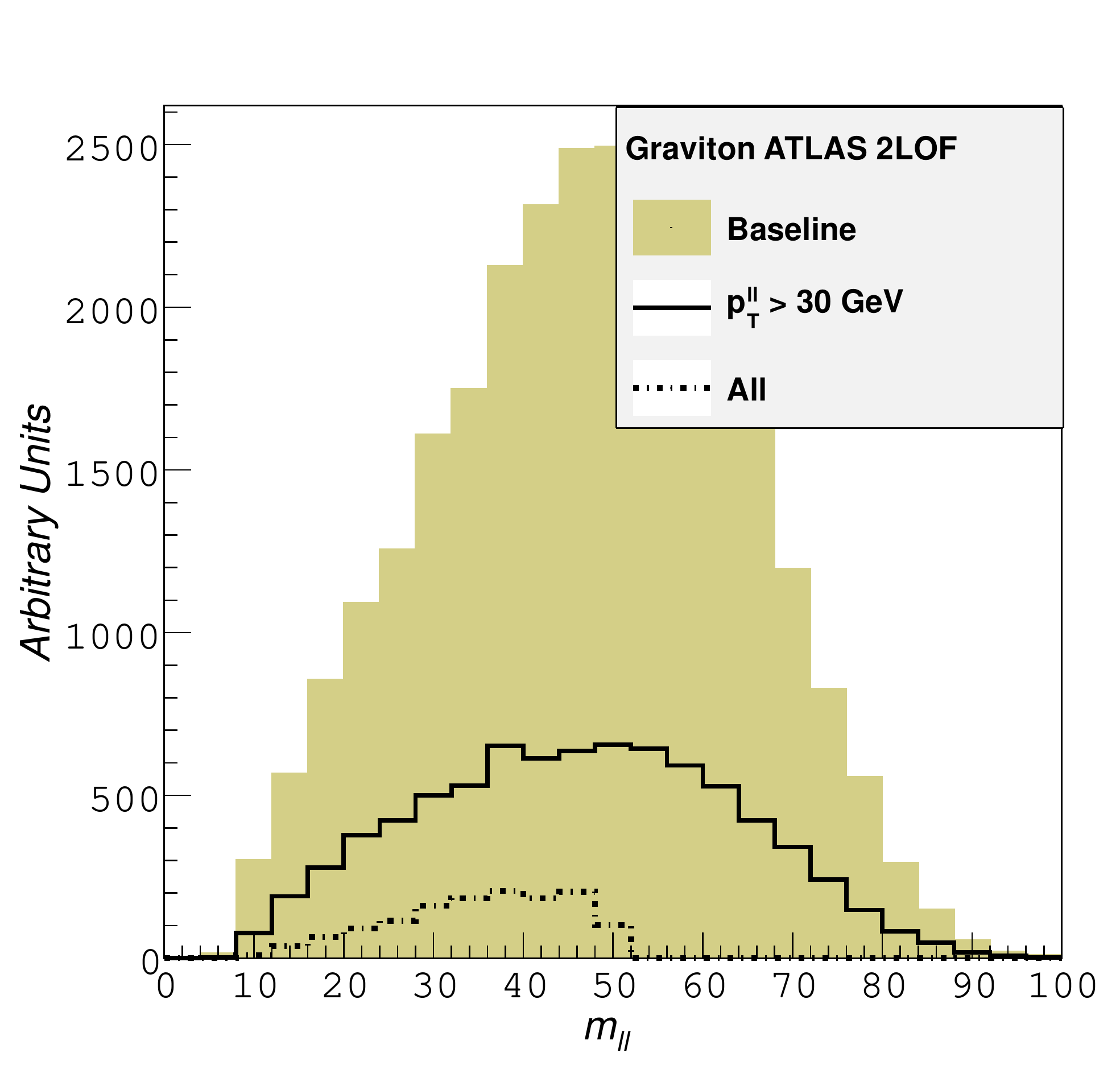}
\includegraphics[scale=0.24]{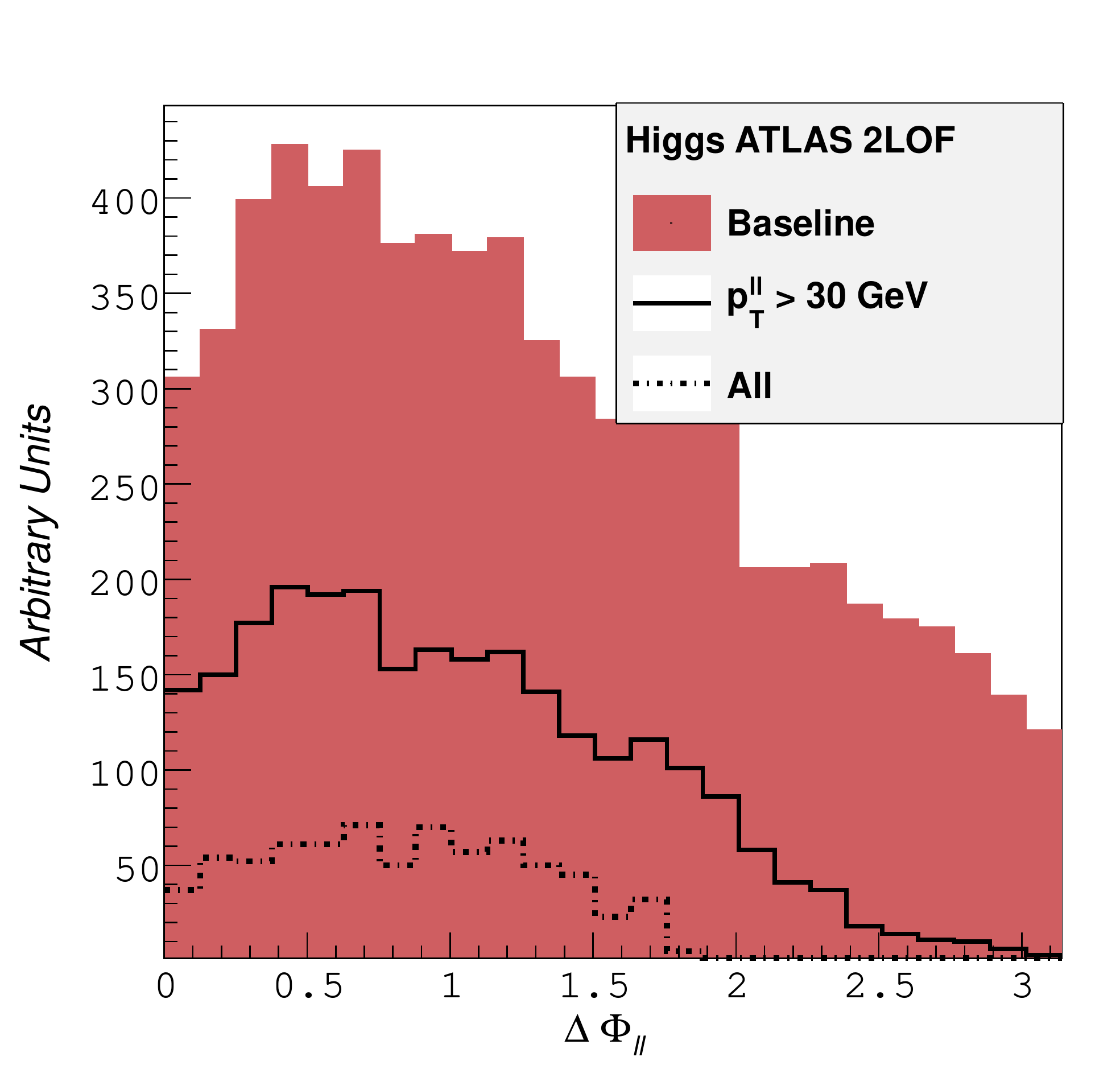}
\includegraphics[scale=0.24]{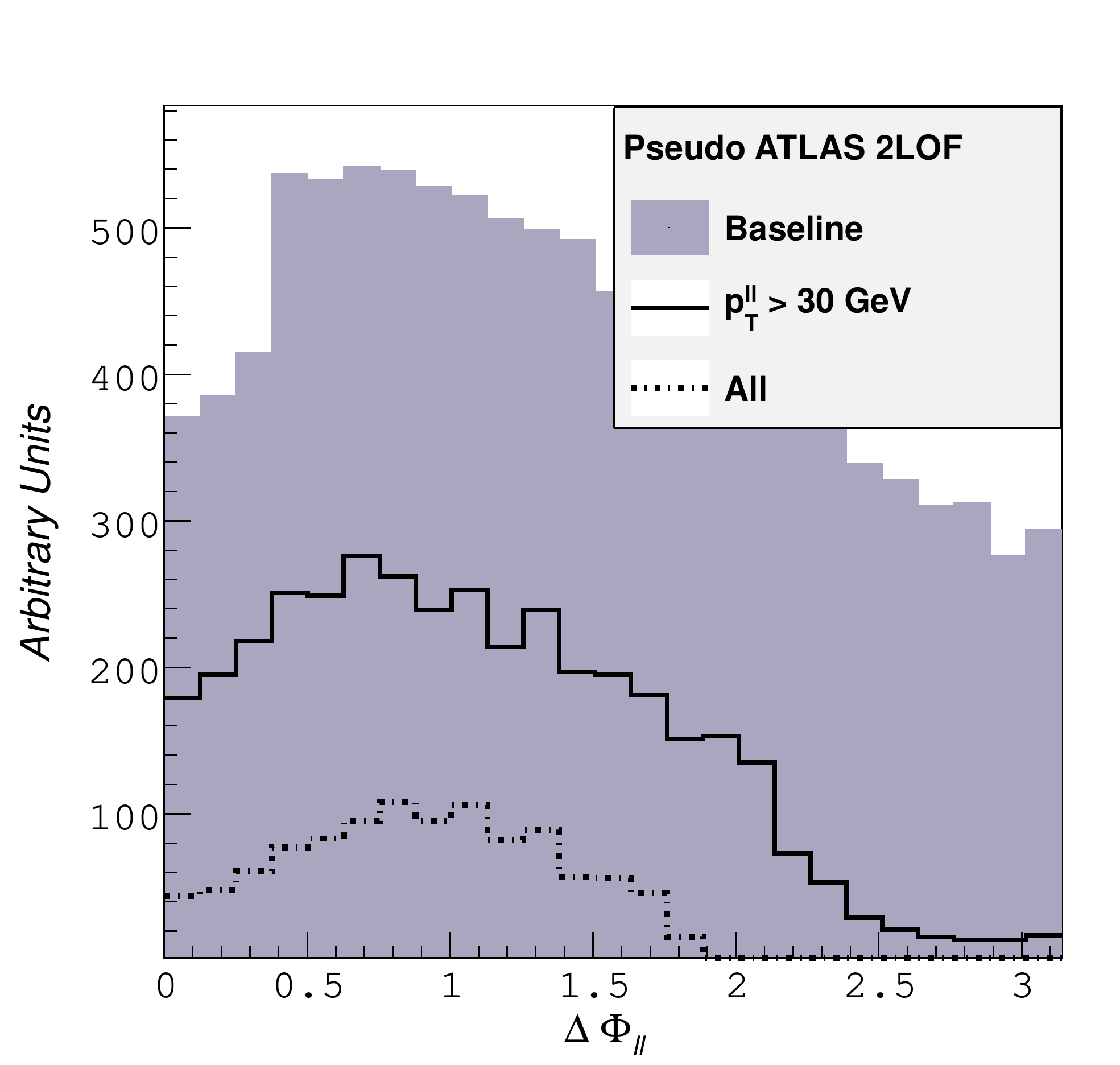}
\includegraphics[scale=0.24]{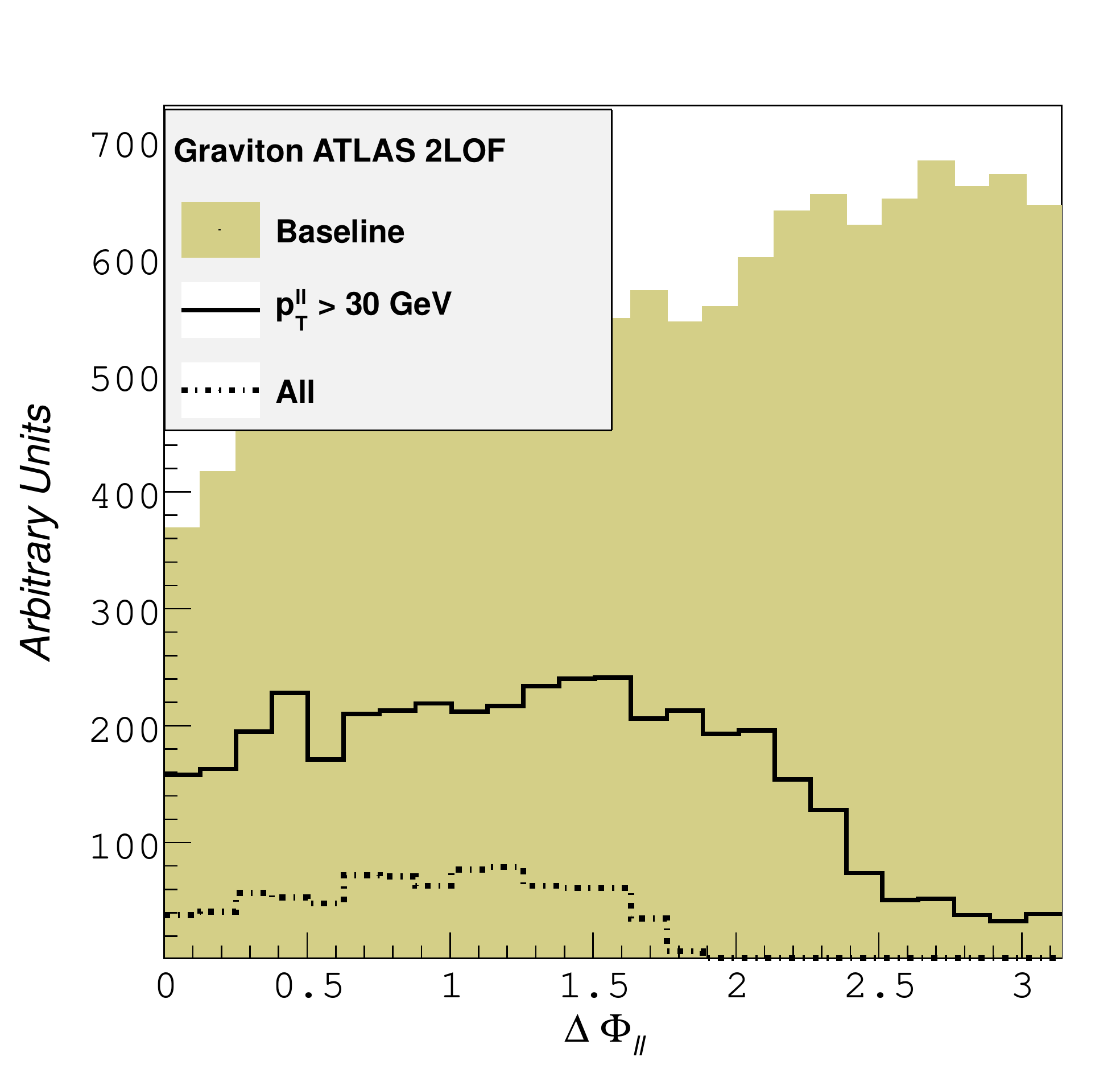}
\caption{\it Results from simulations of the $X_{0^+,0^-,2^+} \to WW^* \to \ell^+\nu_l \ell^-\bar{\nu}_l$ signals,
using {\tt PYTHIA} and {\tt Delphes} and implementing the ATLAS cuts described in the text.
The upper (lower) panels display the $m_{\ell\ell}$ ($\phi_{\ell\ell}$) distributions for $X_{0^+,0^-}$ (left
and centre) and $X_{2^+}$ (right).}
\label{phillATLAS}
\end{figure}

Fig.~\ref{phillATLAS} displays the $m_{\ell\ell}$ (upper panels) and $\phi_{\ell\ell}$
(lower panels) distributions for $X_{0^+, 0^-}$ (left and centre panels) and $X_2$ (right panels) after implementing
in {\tt Delphes} the ATLAS analysis cuts described above. We see that the effect of cuts is dramatic, 
reshaping the distributions in the $X_2$ case so that it resembles the $X_0$ hypothesis. 
This is not only a consequence of the $\Delta \phi_{\ell\ell}$ cut. We have verified that one could loosen or
even remove the $\Delta \phi_{\ell\ell}$ cut: its effect on the background rejection is very mild,
and dropping it would not help to maintain the distinctive kinematic
features of $X_2 \to WW^* \to \ell^+\nu_l \ell^-\bar{\nu}_l$ decay. The initial $p_T^{\ell\ell}$ cut plays a key role
in reducing features due to the anti-parallel preference of the lepton momenta in Fig.~\ref{phillATLAS},
due in turn to the strong correlations between the $\Delta \phi_{\ell\ell}$, $m_{\ell\ell}$  and $p_T^{\ell\ell}$ cuts. 

\begin{figure}[h!]
\centering
\includegraphics[scale=0.45]{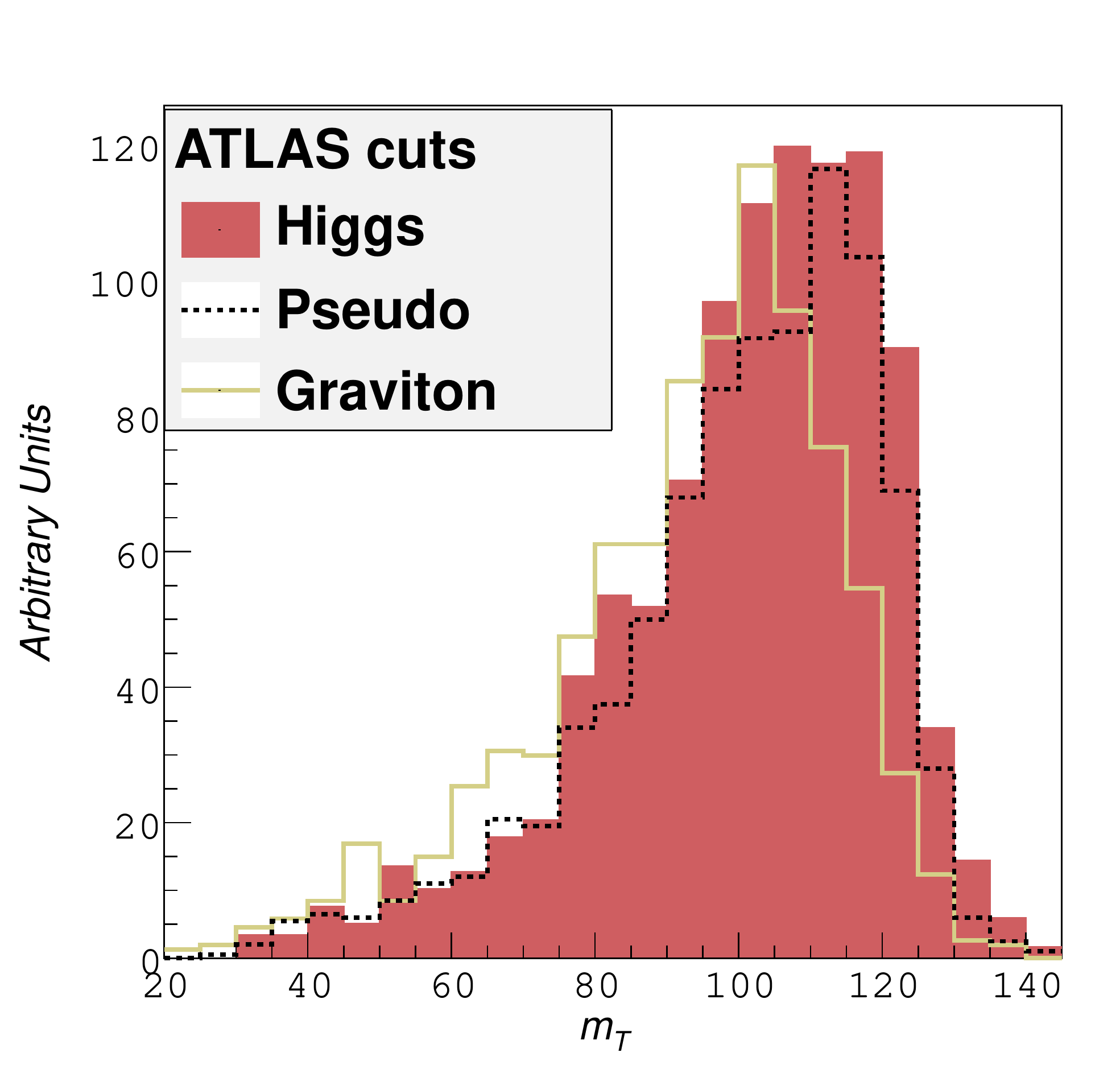}
\caption{\it The transverse mass, $m_T$, distribution after application of all the ATLAS cuts
described in the text, under the three $J^P$ hypotheses.}
\label{mTATLAS}
\end{figure}

In Fig.~\ref{mTATLAS}, we show the $m_T$ distributions found after implementing
all the ATLAS cuts described above, under the $X_{0^+, 0^-}$ and 
graviton-like $X_{2^+}$ hypotheses. They exhibit somewhat different behaviours
within the selected kinematic range. The peaking of the $X_{2^+}$ histogram at
slightly lower $m_T$ than those for $X_{0^+, 0^-}$ is a relic of the differences
in the kinematic distributions before the cuts. Since the neutrino and and antineutrino
are emitted antiparallel in $X_{2^+}$ case, the ${\bar \nu} \nu$ invariant mass has a slight tendency
to be larger than in the $X_{0^+, 0^-}$ cases, implying that $m_T$ tends to fall
further below the $X$ mass of $\sim 125$~GeV.

We have also simulated the corresponding CMS search for $X_{0,2} \to WW^* \to \ell^+\nu_l \ell^-\bar{\nu}_l$. 
The CMS cuts are very similar to those applied by ATLAS, except that CMS requires 
$p_T^{\ell\ell}> $ 45 GeV, $m_{\ell\ell }\in [12,45]$ GeV, $\Delta \Phi_{\ell\ell}< 1.6$ and $m_{T}\in [80,125]$ GeV.
The resulting histograms of $m_{\ell\ell}$ and $\phi_{\ell\ell}$ are very similar to those for ATLAS shown in
Fig.~\ref{phillATLAS}, so we do not show the CMS equivalents.

We display in Table~\ref{cutflow} the ATLAS and CMS cut flows for the different $X_{J^p}$ hypotheses. These
results are based on simulations of 10,000 $0^+$ events, 20,000 $0^-$ events and 30,000 $2^+$ events,
so the statistical errors are negligible. Although there are differences between the numbers of events
surviving different stages in the ATLAS and CMS selection, the end results after applying all the cuts are
very similar. Specifically, we find that the efficiencies for the $0^+$ hypothesis are larger than
those for the $2^+$ hypothesis by factors of 1.86 (1.94) for the ATLAS (CMS) event selections, with the
efficiencies for a pseudoscalar $X_{0^-}$ being about 10\% lower than that for $X_{0^+}$ in both cases.

\begin{table}[t] 
\setlength{\tabcolsep}{5pt}
\center
\begin{tabular}{|c||c|c|c|} 
\hline
Cuts & Scalar $0^+$ & Pseudoscalar $0^-$ & Graviton-like $2^+$
\\
\hline\hline
$p_{T}^{\ell\ell} > $ 30 (45) GeV  & 86 (63) & 83 (58) & 70 (43)\\\hline 
$m^{\ell\ell} < $ 50 (45) GeV  & 66 (49) &  62 (44) & 40  (26)\\\hline 
$\Delta \Phi_{\ell\ell} < $ 1.8 (1.6)  & 63 (49) & 60 (44) & 38 (26)\\\hline 
$m_T \in [93.75,125] ([80,125])$ GeV & 44 (40) &  41 (36) & 22 (21) \\ 
\hline \hline
\end{tabular}
\caption{\it The cutflow evolutions for the ATLAS (CMS) cuts under the three $J^P$
hypotheses. The numbers shown are the cumulative efficiencies to pass the cuts (in percent).}
\label{cutflow} \vspace{-0.35cm}
\end{table}

\subsection{Data analysis under different $J^P$ hypotheses}

We now discuss how this efficiency difference in the experimental selection cuts may in principle be used
to help discriminate between the scalar and graviton-like spin two hypotheses, 
analyze the sensitivity offered by  the current data and estimate the likely sensitivity of the full 2012 data set. 

We parametrize the rescaling of $X_{0^+}$ particle couplings to the $W, Z$ gauge bosons
relative to the Standard Model values by $a_W$ and $a_Z$, respectively, and infer $\lambda_{WZ} \equiv a_W/a_Z$ from the 
measured ratio of the signals in the $WW^\ast$ and $ZZ^\ast$ channels, which is given by
\begin{equation}
R_{WZ} \; = \; \frac{N_W}{N_Z} \; = \; \frac{BR_{SM}(X_{0^+} \to WW^\ast)}{BR_{SM}(X_{0^+} \to ZZ^\ast)}
\left(\frac{a_W}{a_Z}\right)^2\left(\frac{\epsilon_W}{\epsilon_Z}\right) \, ,
\label{eq:Rx}
\end{equation}
where $\epsilon_{W,Z}$ are the efficiencies for the $X_{0^+} \to WW^\ast, ZZ^\ast$ experimental selections.
Likewise, rescaling by $a_{W_2}$ and $a_{Z_2}$ the $W$ and $Z$ couplings of an $X_{2^+}$ particle relative to reference values with
custodial symmetry, one has in an obvious notation
\begin{equation}
R_{WZ_2} \; = \; \frac{N_W}{N_Z} \; = \; \frac{BR_{2^+}(X_{2^+} \to WW^\ast)}{BR_{2^+}(X_{2^+} \to ZZ^\ast)}
\left(\frac{a_{W_2}}{a_{Z_2}}\right)^2\left(\frac{\epsilon_{W_2}}{\epsilon_{Z_2}}\right) \, ,
\label{eq:Rx2}
\end{equation}
which can be used to infer $\lambda_{WZ_2} \equiv a_{W_2}/a_{Z_2}$ in the same way.
Since we use $X \to Z Z^* \to 4 \ell^\pm$ event selections that use only the individual $\ell^\pm$ momenta
(specifically, we do not use the CMS MELA analysis), we may assume that $\epsilon_{Z_2}/\epsilon_Z = 1$.
The value of $\lambda_{WZ_2}$ inferred from the data therefore differs from that of $\lambda_{WZ}$ by the following factor
\begin{equation}
\lambda_{WZ_2} \; = \; \lambda_{WZ} \times \sqrt{ \left(\frac{\epsilon_W}{\epsilon_{W_2}}\right)
\left( \frac{BR_{SM}(X_{0^+} \to WW^\ast)}{BR_{SM}(X_{0^+} \to ZZ^\ast)} {/}  
\frac{BR_{2^+}(X_{2^+} \to WW^\ast)}{BR_{2^+}(X_{2^+} \to ZZ^\ast)} \right)} \, ,
\label{eq:lambda2}
\end{equation}
where the value $\epsilon_{W}/\epsilon_{W_2} \simeq 1.9$ was calculated in the previous Section.

The ratio of the ratio of $X_{0^+,2^+} \to WW^\ast$ and $ZZ^\ast$ branching ratios is not simply unity,
because of the non-trivial dependences of the partial decay widths $\Gamma(X_{0^+,2^+} \to V V^\ast)$
on the masses of the vector bosons $V$. We have used {\tt Madgraph5 v1.4} and  {\tt v1.5} to
calculate the decay widths, and have checked our results against the Standard Model predictions
for $H \to WW^\ast$ and $ZZ^\ast$, and also tested them in the limiting cases of heavy graviton and Higgs, 
when the vector bosons are produced on-shell. In the case of the physical $X$ mass, we find
\begin{equation}
\frac{BR_{2^+}(X_{2^+} \to WW^\ast)}{BR_{2^+}(X_{2^+} \to ZZ^\ast)}/
\frac{BR_{SM}(X_{0^+} \to WW^\ast)}{BR_{SM}(X_{0^+} \to ZZ^\ast)} \; = \; 1.3 \, ,
\label{eq:BR2BR}
\end{equation}
and hence
\begin{equation}
\lambda_{WZ_2} \; = \; \lambda_{WZ} \times 1.2  \, ,
\label{eq:lambdas}
\end{equation}
when $a_W$ and $a_Z$, and hence $\lambda_{WZ}$ and $\lambda_{WZ_2}$,
are extracted using only the $X \to WW^\ast$ and $ZZ^\ast$ inclusive search channels. 

\begin{figure}
\vskip 0.5in
\vspace*{-0.75in}
\begin{minipage}{8in}
\hspace*{-0.7in}
\centerline{
{\includegraphics[height=6cm]{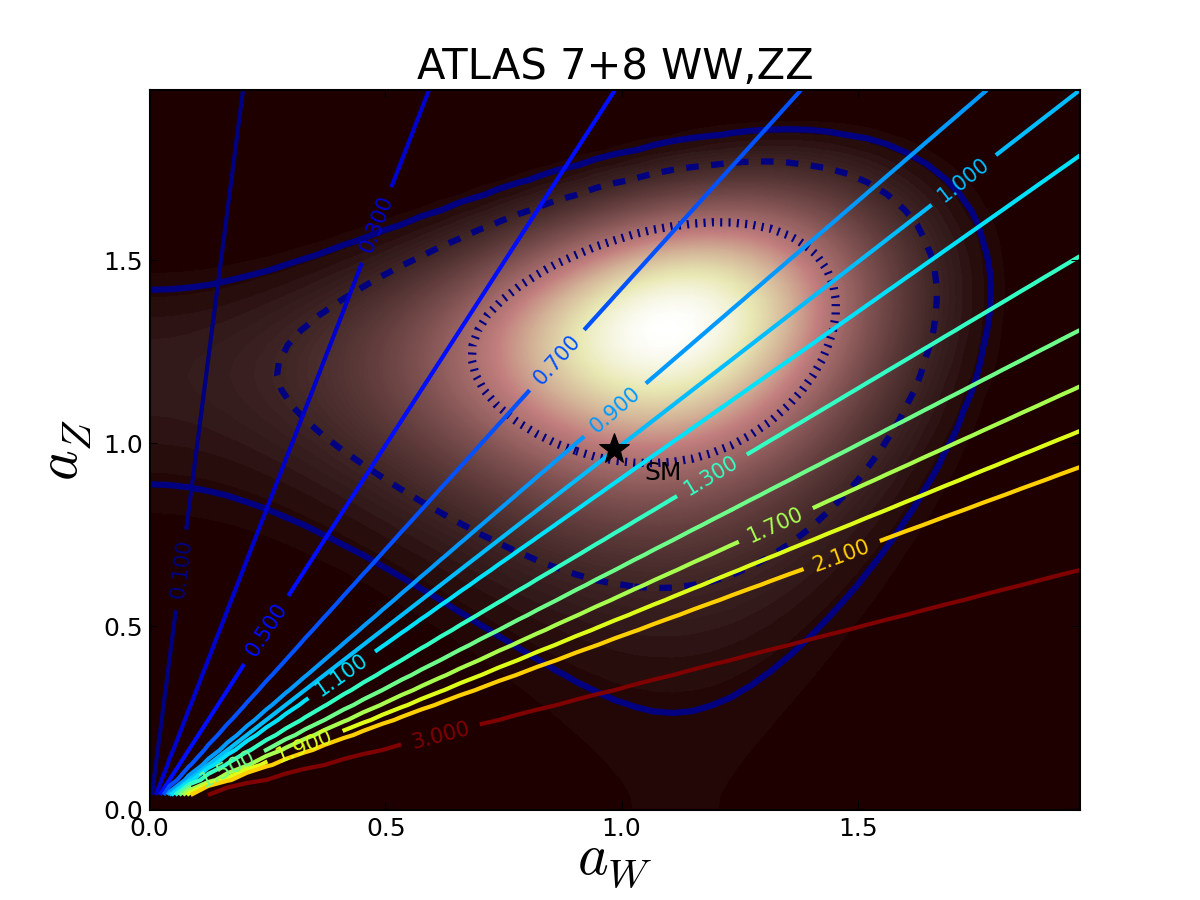}}
{\includegraphics[height=6cm]{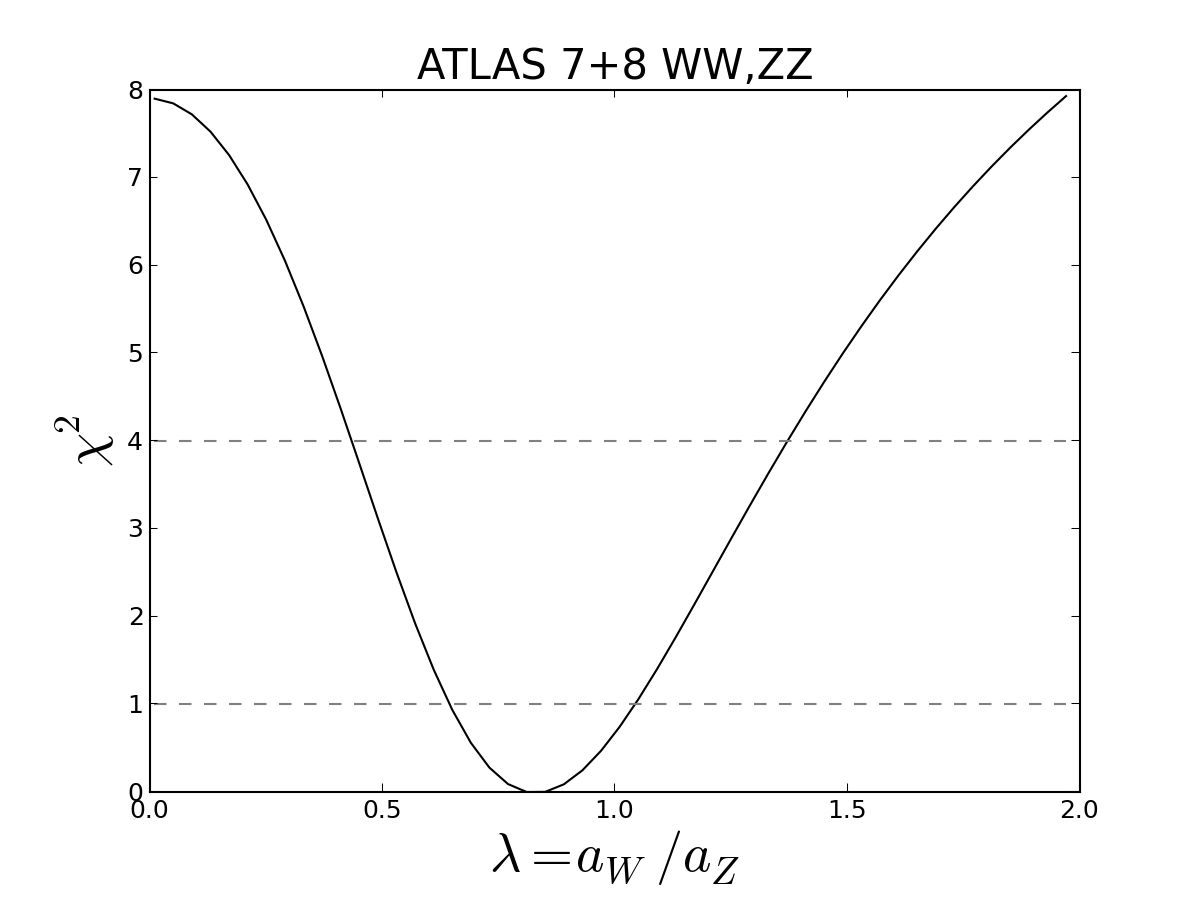}}
}
\hfill
\end{minipage}
\caption{
{\it
Left: The $(a_W, a_Z)$ plane displaying the 68 and 95 \% CL contours
(broken black and solid blue, respectively) obtained from a fit
to the ATLAS 7 and 8~TeV data, with rays corresponding to contours of
$a_W/a_Z$: the Standard Model point $a_W = a_Z = 1$ is indicated by a black star. 
Right: The $\Delta \chi^2$ function relative to the best-fit value in our
ATLAS fit, marginalized over the magnitudes of $a_{W,Z}$, with dashed
horizontal lines at $\Delta \chi^2 = 1, 4$.  
}}  
\label{fig:ATLAS} 
\end{figure}

In Fig.~\ref{fig:ATLAS} we display in the left panel the $(a_W, a_Z)$ plane with
the two-dimensional CL contours that we find in our analysis of the ATLAS 7 and 8~TeV
data combined~\footnote{We cannot use for this purpose the value of $\lambda_{WZ}$ quoted in~\cite{ATLAScouplings},
because this incorporates information from a combination of channels including $X \to \gamma \gamma$,
which supplies information on $a_W$ that  does not apply to the spin-two case.}~\cite{EY2}.
Also shown to guide the eye, in this and subsequent similar plots, are 
rays corresponding to various values of the ratio $a_W/a_Z$. The ray $a_W/a_Z = 1$
passes through the Standard Model point $a_W = a_Z = 1$, which is indicated by a
black star. This point also lies just within the 68\% CL contour, shown as a broken black
line (the 95\% CL contour is a solid blue line). The right panel of Fig.~\ref{fig:ATLAS}
displays the $\Delta \chi^2$ function relative to the best-fit value in our analysis of the
ATLAS 7 and 8~TeV data, marginalized over the magnitudes of $a_{W,Z}$. The combined ATLAS
data do not exhibit any strong preference between the $J^P = 0^+$ and $2^+$ hypotheses,
which correspond to $a_W/a_Z = 1$ and $1/\sqrt{2}$, respectively.

\begin{figure}
\vskip 0.5in
\vspace*{-0.75in}
\begin{minipage}{8in}
\hspace*{-0.7in}
\centerline{
{\includegraphics[height=6cm]{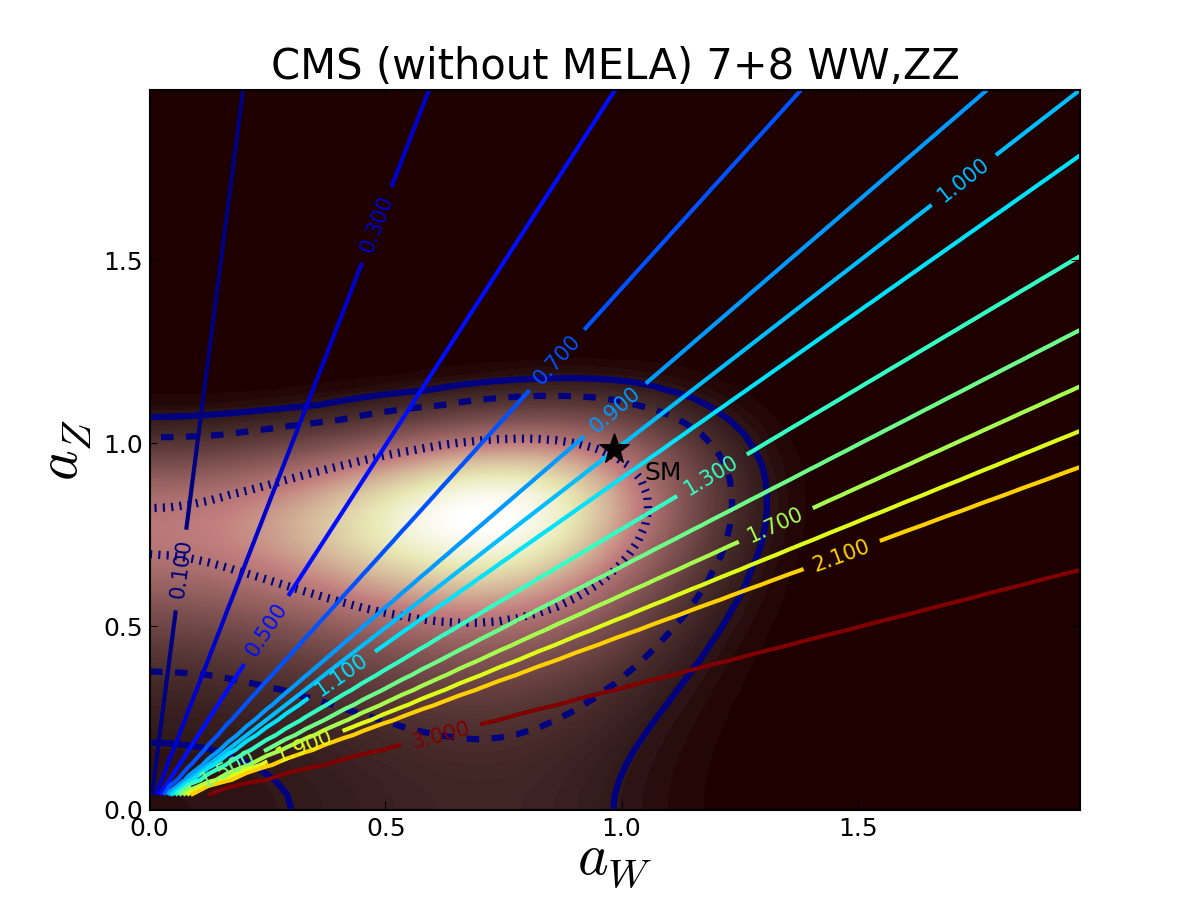}}
{\includegraphics[height=6cm]{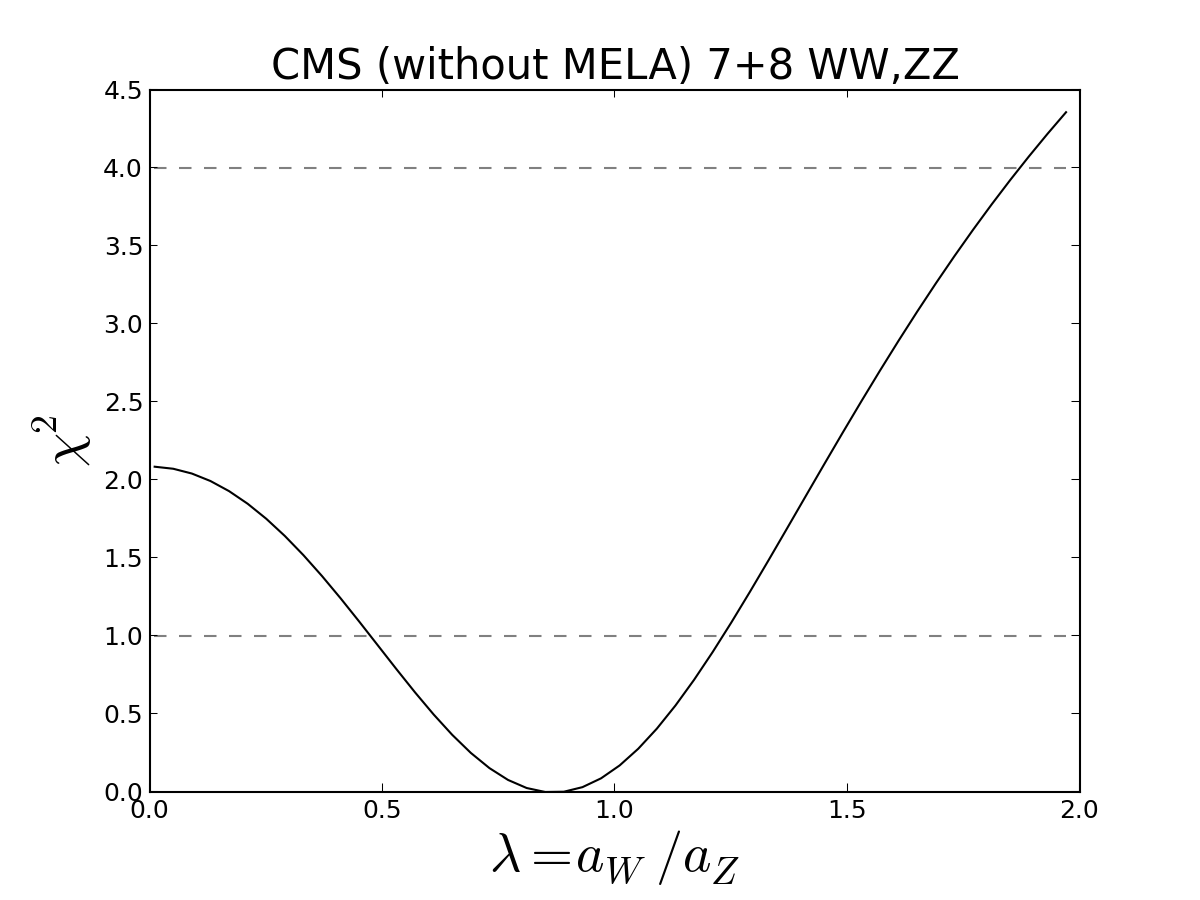}}
}
\hfill
\end{minipage}
\caption{
{\it
As in Fig.~\ref{fig:ATLAS}, but based on our fit to the CMS 7 and 8~TeV data,
which does not include the final MELA selection made in~\cite{CMSICHEP2012}.
}}  
\label{fig:CMS} 
\end{figure}

Fig.~\ref{fig:CMS} displays a similar pair of panels for our analysis of the available
CMS 7 and 8~TeV data. We note here that we do not use the final
CMS result for $X \to W W^\ast$ signal, which include an MELA selection
that we do not model. Instead, we use the expected signal, background and observed event numbers
shown in Table~3 of~\cite{CMSICHEP2012}, which correspond directly to the CMS event selection
and efficiency found in our simulation above. As in the ATLAS case above, the CMS
data also do not exhibit a strong preference between the $J^P = 0^+$ and $2^+$ hypotheses.

\begin{figure}
\vskip 0.5in
\begin{minipage}{8in}
\hspace*{-0.7in}
\centerline{{\includegraphics[height=6cm]{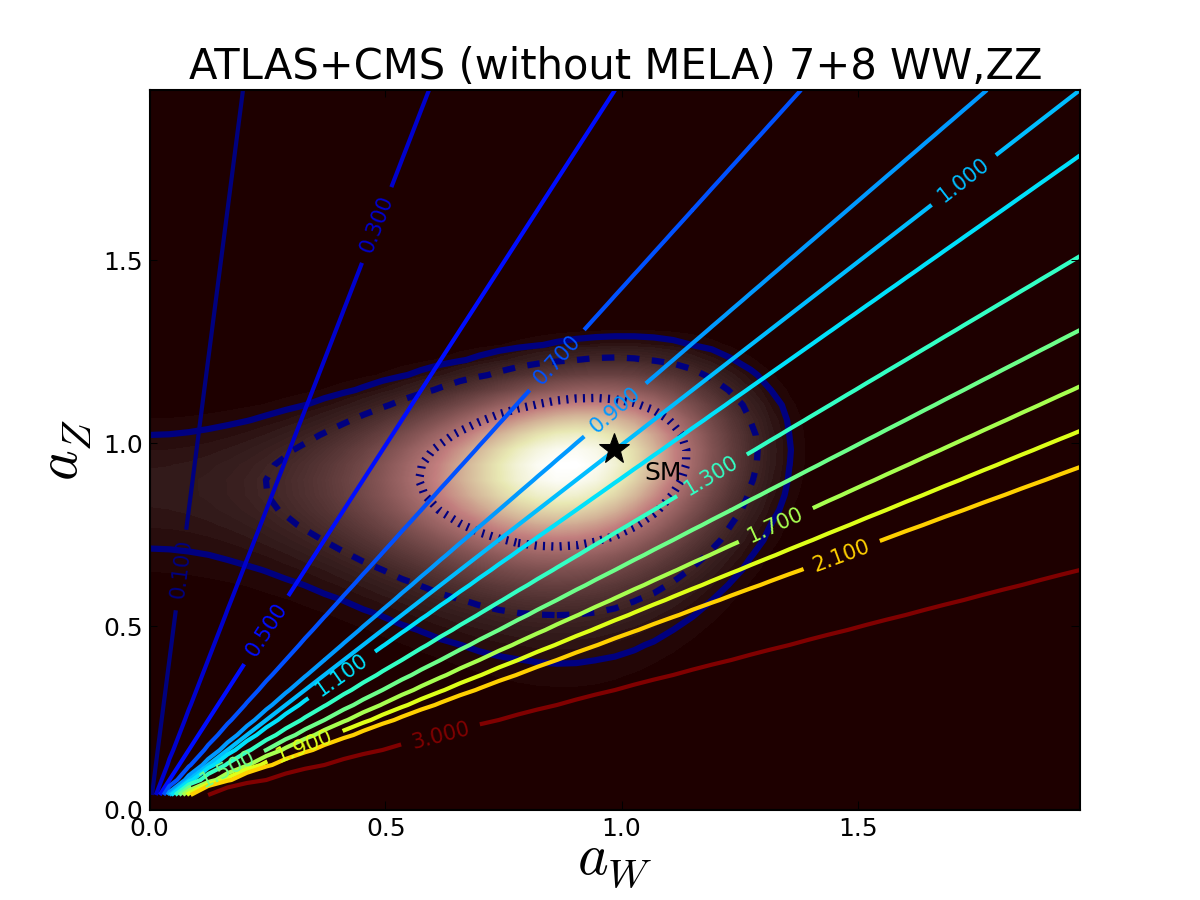}}
{\includegraphics[height=6cm]{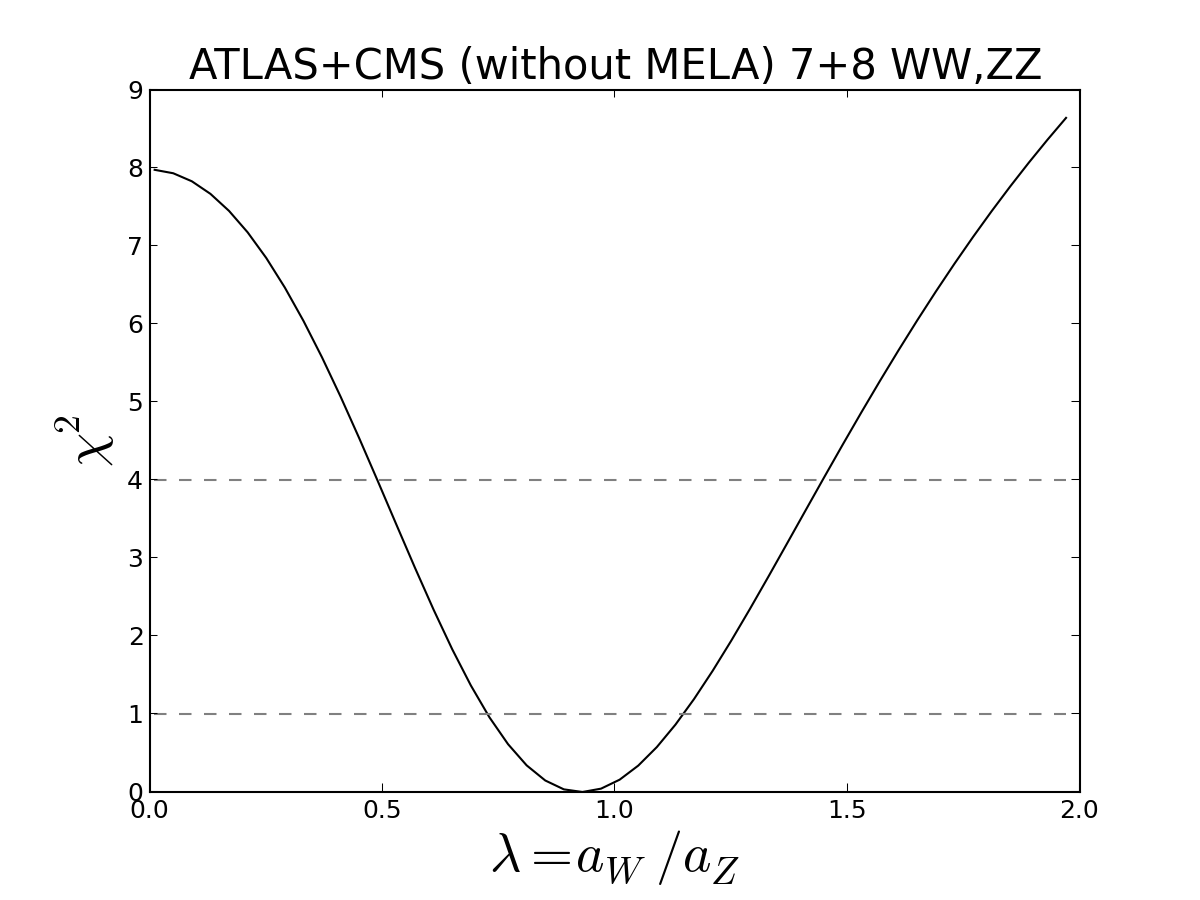}}}
\hfill
\end{minipage}
\caption{
{\it
As in Fig.~\ref{fig:ATLAS}, but based on our combined fit to the ATLAS and CMS 7 and 8~TeV data.
}}  
\label{fig:combined} 
\end{figure}

Fig.~\ref{fig:combined} displays a similar pair of panels for our combined analysis of the available
ATLAS and CMS 7 and 8~TeV data. The data sets in combination provide no significant
discrimination between the $J^P = 0^+$ and $2^+$ hypotheses. Our combined $\chi^2$ analysis yields
\begin{equation}
\lambda_{WZ} \; \equiv \; \frac{a_W}{a_Z} \; = \; 0.93^{+ 0.24}_{- 0.20} \, .
\label{combinedchi2lambda}
\end{equation}
Using (\ref{eq:lambdas}), the corresponding value in the $X_{2^+}$ case is
\begin{equation}
\lambda_{WZ_2} \; = \; 1.12^{+ 0.29}_{- 0.24} \, ,
\label{combinedchi2lambda2}
\end{equation}
and we see that both results are compatible with unity.

This result is based on $\sim 5$/fb of data at each of 7 and 8 TeV analyzed by each of ATLAS
and CMS. At the time of writing, each experiment has now recorded $\sim 15$/fb of data at
8 TeV, corresponding approximately to a doubling of the statistics and potentially to a reduction
in the uncertainty in $a_W/a_Z$ by a factor $\sim 1.4$. It is anticipated that each experiment
might record $\sim 30$/fb of data by the end of 2012, corresponding to reductions in the
statistical errors in (\ref{combinedchi2lambda}, \ref{combinedchi2lambda2})
by factors $\sim \sqrt{3}$. If the central value remained the same, (\ref{combinedchi2lambda2})
would become $\lambda_{WZ_2} = 1.12^{+ 0.17}_{- 0.14}$, whereas if the central value
were to correspond to $\lambda_{WZ} = 1$, one would have $\lambda_{WZ_2} = 1.21^{+ 0.17}_{- 0.14}$.
We conclude that the best one could reasonably hope for with the 2012 data would be a deviation from the
custodial symmetry prediction $\lambda_{WZ_2} = 1$ of about one $\sigma$.

\section{Overview and Prospects}

We have explored in this paper two strategies to help determine the spin of the new particle $X$
discovered recently by the ATLAS and CMS Collaborations. One of these exploits 
the angular distribution of the final-state photons in
$gg \to X \to \gamma \gamma$ decay, and the other exploits the angular
distributions and correlations in $X \to W W^* \to \ell^+ \ell^- \nu {\bar \nu}$ decay, which are in
principle quite different for different spin assignments $J^P = 0^+, 0^-$ and $2^+$ for the $X$
particle. 

We have shown how the 2012 LHC data could be used to study the
angular distribution in $gg \to X \to \gamma \gamma$ decay and provide potentially
significant discrimination between the spin-zero and spin-two hypotheses. A simple angular
asymmetry measurement gives a discrimination power approaching that possible with a
full LLR analysis. We include the effects of backgrounds in samples with both high and low
$S/B = 0.42$ and 0.19, corresponding in spirit to CMS event categories. We find that the
present data should already provide some discrimination between the
$J^P = 0^\pm$ and $2^+$ hypotheses, and that an analysis of the full 2012 data set
could provide a separation significance of $\sim 3 \, \sigma$ if a conservative
symmetric interpretation of the LLR statistic is used, rising to above $6 \, \sigma$ if a less
conservative asymmetric interpretation is used.

We have analyzed the sensitivities of the published results of the
ATLAS and CMS searches for $H \to W W^*$ to the spin-parity of the $X$ particle. Simulating these searches
using {\tt PYTHIA} and {\tt Delphes}, we have shown that the ATLAS and CMS
experimental selections suppress the kinematic differences between $0^+$ and $2^+$ decays. Therefore, an analysis based on kinematic shapes would be rather inconclusive based in the published cuts.   

One could hope for retaining some of the kinematic differences by  changing the cuts, but those changes are limited by the need to suppress the large Standard Model $WW$ background. A more hopeful strategy, and the one we developed in this paper, is to use of the approximate custodial symmetry in both spin-zero and -two hypothesis, to relate the $WW^\ast$ and $ZZ^\ast$ channels.
In $WW^\ast$, the efficiencies of the searches differ by a factor $\simeq 1.9$ for $X_{0^+}$ and $X_{2^+}$. On the other
hand, we find that the ratio of $X_{2^+} \to WW^\ast$ and $ZZ^\ast$
branching ratios is $\simeq 1.3$ larger than the corresponding ratio of branching ratios in the $0^+$ case.
The current measurements by ATLAS and CMS of the
ratio of experimental rates for $X \to W W^* \to \ell^+ \ell^- \nu {\bar \nu}$ and 
$X \to Z Z^* \to 4 \ell^\pm$ are compatible with custodial symmetry in both the $0^+$ and $2^+$
cases, and we do not expect that this will improve significantly with the full 2012 set of LHC data.

Many other strategies to discriminate between different spin-parity hypotheses for the $X$
particle have been proposed, including kinematic correlations in $X \to Z Z^* \to 4 \ell^\pm$
decay and the threshold behaviour of associated $W/Z + X$ production, as discussed in the Introduction. We have every
confidence that the spin-parity of the $X$ particle will be pinned down within a few months,
in good time for the nominations for the 2013 Nobel Physics Prize.

\section*{Acknowledgements}
We thank Oliver Buchm\"uller, Ben Gripaios, Rakhi Mahbubani, Eduard Mass\'o and Pierre Savard for valuable discussions.
The work of JE was supported partly by the London
Centre for Terauniverse Studies (LCTS), using funding from the European
Research Council via the Advanced Investigator Grant 267352.
The work of DSH
was supported partly by the Korea Foundation for International Cooperation of Science \& Technology
(KICOS) and the Basic Science Research Programme of the National Research Foundation of
Korea (2012-0002959). The work of TY was supported by a Graduate Teaching Assistantship from
King's College London. JE, DSH and VS thank CERN for kind hospitality, and TY thanks
Prof. T. Kobayashi and the Bilateral International Exchange Program of Kyoto University for kind
hospitality.


\begin{thebibliography}{99}

\bibitem{ATLASICHEP2012}
G.~Aad {\it et al.}  [ATLAS Collaboration],
  Phys.\ Lett.\ B {\bf 716} (2012) 1
  [arXiv:1207.7214 [hep-ex]]; see also
  F.~Gianotti, talk on behalf of the
ATLAS Collaboration at CERN, 4th July, 2012,\\
{\tt https://cms-docdb.cern.ch/cgi-bin/PublicDocDB//ShowDocument?docid=6126.}


\bibitem{CMSICHEP2012}
S.~Chatrchyan {\it et al.}  [CMS Collaboration],
  Phys.\ Lett.\ B {\bf 716} (2012) 30
  [arXiv:1207.7235 [hep-ex]]; see also
J.~Incandela, talk on behalf of the
CMS Collaboration at CERN, 4th July, 2012,\\
{\tt https://cms-docdb.cern.ch/cgi-bin/PublicDocDB//ShowDocument?docid=6125.}

\bibitem{EY2}
 J.~Ellis and T.~You,
  JHEP {\bf 1209}, 123 (2012)
  [arXiv:1207.1693 [hep-ph]].
   
\bibitem{postdiscovery}
For other analyses of the $X$ particle couplings, see also: 
D.~Carmi, A.~Falkowski, E.~Kuflik and T.~Volanski,
arXiv:1202.3144 [hep-ph];
A.~Azatov, R.~Contino and J.~Galloway,
  JHEP {\bf 1204} (2012) 127  [hep-ph/1202.3415].
J.R.~Espinosa, C.~Grojean, M.~Muhlleitner and M.~Trott,
arXiv:1202.3697 [hep-ph];
P.~P.~Giardino, K.~Kannike, M.~Raidal and A.~Strumia,
arXiv:1203.4254 [hep-ph];
T.~Li, X.~Wan, Y.~Wang and S.~Zhu,
arXiv:1203.5083 [hep-ph];
M.~Rauch,
arXiv:1203.6826 [hep-ph];
 J.~Ellis and T.~You,
JHEP {\bf 1206} (2012) 140,
[arXiv:1204.0464 [hep-ph]];
A.~Azatov, R.~Contino, D.~Del Re, J.~Galloway, M.~Grassi and S.~Rahatlou,
arXiv:1204.4817 [hep-ph];
M.~Klute, R.~Lafaye, T.~Plehn, M.~Rauch and D.~Zerwas,
arXiv:1205.2699 [hep-ph];
J.R.~Espinosa, M.~Muhlleitner, C.~Grojean and M.~Trott,
arXiv:1205.6790 [hep-ph];
 D.~Carmi, A.~Falkowski, E.~Kuflik and T.~Volansky,
 arXiv:1206.4201 [hep-ph]; 
 M.~J.~Dolan, C.~Englert and M.~Spannowsky,
 arXiv:1206.5001 [hep-ph];
 J.~Chang, K.~Cheung, P.~Tseng and T.~Yuan,
 arXiv:1206.5853 [hep-ph];
 S.~Chang, C.~A.~Newby, N.~Raj and C.~Wanotayaroj,
 arXiv:1207.0493 [hep-ph];
 I.~Low, J.~Lykken and G.~Shaughnessy,
 arXiv:1207.1093 [hep-ph];
 T.~Corbett, O.~J.~P.~Eboli, J.~Gonzalez-Fraile and M.~C.~Gonzalez-Garcia,
  arXiv:1207.1344 [hep-ph];
P.~P.~Giardino, K.~Kannike, M.~Raidal and A.~Strumia, 
 arXiv:1207.1347 [hep-ph];
M.~Montull and F.~Riva,
arXiv:1207.1716 [hep-ph];
J.~R.~Espinosa, C.~Grojean, M.~Muhlleitner and M.~Trott,
arXiv:1207.1717 [hep-ph];
D.~Carmi, A.~Falkowski, E.~Kuflik, T.~Volansky and J.~Zupan, 
arXiv:1207.1718 [hep-ph];
S.~Banerjee, S.~Mukhopadhyay and B.~Mukhopadhyaya,
JHEP {bf 10} (2012) 062,
[arXiv:1207.3588 [hep-ph]];
F.~Bonner, T.~Ota, M.~Rauch and W.~Winter,
arXiv:1207.4599 [hep-ph];
T.~Plehn and M.~Rauch,
arXiv:1207.6108 [hep-ph];
A.~Djouadi,
arXiv:1208.3436 [hep-ph];
B.~Batell, S.~Gori and L.~T.~Wang,
arXiv:1209.6832 [hep-ph].
  
\bibitem{TevatronJulySearch}
T.~Aaltonen {\it et al.}  [CDF and D0 Collaborations],
  arXiv:1207.6436 [hep-ex]; see also
  TEVNPH Working Group, for the CDF and D0 Collaborations
arXiv:1207.0449 [hep-ex].

  
\bibitem{EHSY}
  J.~Ellis, D.~S.~Hwang, V.~Sanz and T.~You,
  arXiv:1208.6002 [hep-ph].
  
  \bibitem{zz}
See, for example
S.~Y.~Choi, D.~J.~.~Miller, M.~M.~Muhlleitner and P.~M.~Zerwas,
  Phys.\ Lett.\  B {\bf 553} (2003) 61
  [arXiv:hep-ph/0210077];
K.~Odagiri,
  JHEP {\bf 0303} (2003) 009
  [arXiv:hep-ph/0212215];
C.~P.~Buszello, I.~Fleck, P.~Marquard and J.~J.~van der Bij,
  Eur.\ Phys.\ J.\  C {\bf 32} (2004) 209
  [arXiv:hep-ph/0212396];
A.~Djouadi,
  Phys.\ Rept.\  {\bf 457} (2008) 1
  [arXiv:hep-ph/0503172];
  C.~P.~Buszello and P.~Marquard,
  arXiv:hep-ph/0603209;
A.~Bredenstein, A.~Denner, S.~Dittmaier and M.~M.~Weber,
  Phys.\ Rev.\  D {\bf 74} (2006) 013004
  [arXiv:hep-ph/0604011];
 P.~S.~Bhupal Dev, A.~Djouadi, R.~M.~Godbole, M.~M.~Muhlleitner and S.~D.~Rindani,
  Phys.\ Rev.\ Lett.\  {\bf 100} (2008) 051801
  [arXiv:0707.2878 [hep-ph]];
R.~M.~Godbole, D.~J.~.~Miller and M.~M.~Muhlleitner,
  JHEP {\bf 0712} (2007) 031
  [arXiv:0708.0458 [hep-ph]];
K.~Hagiwara, Q.~Li and K.~Mawatari,
  JHEP {\bf 0907} (2009) 101
  [arXiv:0905.4314 [hep-ph]];
A.~De Rujula, J.~Lykken, M.~Pierini, C.~Rogan and M.~Spiropulu,
  Phys.\ Rev.\  D {\bf 82} (2010) 013003
  [arXiv:1001.5300 [hep-ph]];
 C.~Englert, C.~Hackstein and M.~Spannowsky,
  Phys.\ Rev.\  D {\bf 82} (2010) 114024
  [arXiv:1010.0676 [hep-ph]];
  U.~De Sanctis, M.~Fabbrichesi and A.~Tonero,
  Phys.\ Rev.\  D {\bf 84} (2011) 015013
  [arXiv:1103.1973 [hep-ph]];
  V.~Barger and P.~Huang,
  Phys.\ Rev.\ D {\bf 84} (2011) 093001
  [arXiv:1107.4131 [hep-ph]];
  S.~Bolognesi, Y.~Gao, A.~V.~Gritsan, K.~Melnikov, M.~Schulze, N.~V.~Tran and A.~Whitbeck,
  arXiv:1208.4018 [hep-ph];
  R.~Boughezal, T.~J.~LeCompte and F.~Petriello,
  arXiv:1208.4311 [hep-ph];
  D.~Stolarski and R.~Vega-Morales,
  arXiv:1208.4840 [hep-ph];
S.~Y.~Choi, M.~M.~Muhlleitner and P.~M.~Zerwas,
arXiv:1209.5268 [hep-ph].

  \bibitem{us-G}  
  R.~Fok, C.~Guimaraes, R.~Lewis and V.~Sanz,
  arXiv:1203.2917 [hep-ph].
 
\bibitem{RSbulk}
T.~Gherghetta and A.~Pomarol,
  Nucl.\ Phys.\ B {\bf 586}, 141 (2000)
  [hep-ph/0003129].
   T.~Gherghetta and A.~Pomarol,
  Nucl.\ Phys.\ B {\bf 602}, 3 (2001)
  [hep-ph/0012378].
  
\bibitem{Randall:2002tg} 
  L.~Randall, V.~Sanz and M.~D.~Schwartz,
  JHEP {\bf 0206}, 008 (2002)
  [hep-th/0204038].

 \bibitem{gap-metrics}
   J.~Hirn and V.~Sanz,
  Phys.\ Rev.\ D {\bf 76}, 044022 (2007)
  [hep-ph/0702005 [HEP-PH]].
  Phys.\ Rev.\ Lett.\  {\bf 97}, 121803 (2006)
  [hep-ph/0606086].
  JHEP {\bf 0703}, 100 (2007)
  [hep-ph/0612239].
 
  \bibitem{pdg}
   S.~Eidelman {\it et al.}  [Particle Data Group Collaboration],
  Phys.\ Lett.\ B {\bf 592}, 1 (2004).
  
\bibitem{lisa-liam}
 A.~L.~Fitzpatrick, J.~Kaplan, L.~Randall and L.~-T.~Wang,
  JHEP {\bf 0709}, 013 (2007)
  [hep-ph/0701150].
  
  \bibitem{Grosmann-ferm} 
  Y.~Grossman and M.~Neubert,
  Phys.\ Lett.\ B {\bf 474}, 361 (2000)
  [hep-ph/9912408].

\bibitem{Gao} 
  Y.~Gao, A.~V.~Gritsan, Z.~Guo, K.~Melnikov, M.~Schulze and N.~V.~Tran,
  Phys.\ Rev.\ D {\bf 81} (2010) 075022
  [arXiv:1001.3396 [hep-ph]].
 
\bibitem{DaeSung}
J.~Ellis and D.~S.~Hwang,
  JHEP {\bf 1209} (2012) 071
[arXiv:1202.6660 [hep-ph]].
  
 \bibitem{MG5} 
  J.~Alwall {\it et al.},
{\it MadGraph 5 : Going Beyond,}
  JHEP {\bf 1106}, 128 (2011)
  [arXiv:1106.0522 [hep-ph]].
  
\bibitem{PYTHIA} 
  T.~Sjostrand, S.~Mrenna and P.~Z.~Skands,
  {\it PYTHIA 6.4 Physics and Manual,}
  JHEP {\bf 0605}, 026 (2006)
  [hep-ph/0603175].
  
\bibitem{Delphes} 
  S.~Ovyn, X.~Rouby and V.~Lemaitre,
  {\it Delphes, a framework for fast simulation of a generic collider experiment,}
  arXiv:0903.2225 [hep-ph].
 
\bibitem{Alves} 
 A.~Alves,
  arXiv:1209.1037 [hep-ph].

\bibitem{Cousins} 
R.~Cousins, J.~Mumford, J.~Tucker, and V.~Valuev,
JHEP {\bf 11} (2005) 046.


 \bibitem{Feynrules}
 N.~D.~Christensen and C.~Duhr,
 {\it FeynRules - Feynman rules made easy,}
  Comput.\ Phys.\ Commun.\  {\bf 180} (2009) 1614
  [arXiv:0806.4194 [hep-ph]].
  
\bibitem{UFO}
 C.~Degrande, C.~Duhr, B.~Fuks, D.~Grellscheid, O.~Mattelaer and T.~Reiter,
{\it UFO - The Universal FeynRules Output,}
  Comput.\ Phys.\ Commun.\  {\bf 183} (2012) 1201
  [arXiv:1108.2040 [hep-ph]].

\bibitem{ATLAScouplings}
ATLAS Collaboration,\\
{\tt https://atlas.web.cern.ch/Atlas/GROUPS/PHYSICS/CONFNOTES/ATLAS-CONF-2012-127/}.

\end{thebibliography}
\end{document}